\documentclass{article}

\usepackage[utf8]{inputenc}%
\usepackage[T1]{fontenc}%
\usepackage[margin=1in]{geometry}%
\usepackage[authoryear]{natbib}%
\usepackage{authblk}%
\usepackage{rotating}%
\usepackage{graphicx}%
\usepackage{adjustbox}%
\usepackage{multirow}%
\usepackage{amsmath,amssymb,amsfonts}%
\usepackage{amsthm}%
\usepackage{mathrsfs}%
\usepackage[title]{appendix}%
\usepackage{xcolor}%
\usepackage{textcomp}%
\usepackage{manyfoot}%
\usepackage{booktabs}%
\usepackage{algorithm}%
\usepackage{algorithmicx}%
\usepackage{algpseudocode}%
\usepackage{listings}%
\usepackage{dsfont}%
\usepackage{hyperref}%

\newcommand{\eps}{\varepsilon}
\providecommand{\permil}{\text{\textperthousand}}
\newcommand{\E}{\mathds{E}}
\newcommand{\C}{\mathds{C}}
\newcommand{\R}{\mathds{R}}
\newcommand{\cov}{\,\mbox{cov}}
\begin{document}

\title{Continuous-time state space methods for $\delta^{18}$O and $\delta^{13}$C in the Cenozoic Era}


\author[1,2]{Mikkel Bennedsen}

\author[1,2]{Eric Hillebrand\thanks{Corresponding author: \texttt{ehillebrand@econ.au.dk}}}

\author[3]{Siem Jan Koopman}

\author[1,2]{Kathrine Larsen}

\author[4]{Rachel Lupien}

\affil[1]{Department of Economics and Business Economics, Aarhus University, Fuglesangs Allé 4, 8210 Aarhus, Denmark}

\affil[2]{Center for Research in Energy: Economics and Markets (CoRE), Aarhus University, Fuglesangs Allé 4, 8210 Aarhus, Denmark}

\affil[3]{Department of Econometrics, Vrije Universiteit Amsterdam, De Boelelaan 1105, 1081 HV Amsterdam, The Netherlands}

\affil[4]{Department of Geoscience, Aarhus University, Høegh-Guldbergs Gade 2, 8000 Aarhus, Denmark}

\date{\today}




\maketitle

\begin{abstract}
Time series analysis of $\delta^{18}$O and $\delta^{13}$C from benthic foraminifera for paleoclimatology poses significant challenges. The data span tens of millions of years, with sparse early records, dense later ones, uneven time stamps, and occasional multiples. These time series are largely non-stationary, exhibiting temporary, varying trends. We propose a continuous-time state space framework that handles these irregularities effectively. Univariate signal-plus-noise models are specified for $\delta^{18}$O and $\delta^{13}$C, with parameters estimated via maximum likelihood using Kalman filter recursions for signal extraction and likelihood evaluation. The framework interprets state space models as time-domain Butterworth filters. Measurement-error variances are differentiated by deep-sea drill site, including site-specific level offsets, and the record is partitioned into sub-periods reflecting the distinct climate states that drive the transition variance. Two extensions of the univariate model are explored: \emph{(i)} modifying the signal specification for the Kalman filter to approximate a Butterworth filter of any order, and \emph{(ii)} specifying a bivariate signal-plus-noise model for joint analysis. Results reveal substantial signal changes during the ``icehouse'' period (3.3 to 0.0006 Ma); the correlation between $\delta^{18}$O and $\delta^{13}$C signals is generally positive but turns negative during this period.
\end{abstract}




\section{Introduction}\label{sec1}

Measurements of the deviation in the ratio of the oxygen isotopes $^{18}$O and $^{16}$O in ocean sediment cores containing fossil benthic foraminifera allow for obtaining a time series record covering the Cenozoic, i.e., the last 66 million years \citep{westerhold2020,zachos2001}. Since the ratio of the isotopes depends on prevailing temperatures and the amount of water locked in ice sheets at the time of formation of the shell, these measurements can serve as a proxy for global temperature and continental ice volume. Similarly, deviations in the ratio of the carbon isotopes $^{13}$C and $^{12}$C are used as a proxy for the carbon cycle, which integrates contributions from the terrestrial biosphere, the storage of carbon in the ocean and sediments, and atmospheric CO$_2$.

In this paper, we propose a class of continuous-time state space models for the time series of $\delta^{18}$O and $\delta^{13}$C, the deviations in the ratios from a mean ocean water standard. This class of models addresses the central challenges that the time series pose, such as non-stationarity with trends in changing directions, irregular spacing of the time stamps of the observations, and multiple observations at certain time stamps. Additionally, the model can take into account different studies of origin of the data, varying climate states covered by the data, and covariation in the two series, allowing one to inform the estimation of a model for the other. The models are specified in the time domain, and their parameters are estimated by maximum likelihood. The models have a direct interpretation as Butterworth filters in the frequency domain \citep{gomez2001,harvey2003}. Their theory is well developed, see, for example, \cite{Harvey1989} and \cite{Durbin2012} for textbook treatments, and they can be estimated in relatively short time on laptop computers.

The data set employed in this paper is provided in the study by \cite{westerhold2020}, where $\delta^{18}$O and $\delta^{13}$C measurements from 34 different studies are collated and harmonized in a single data file. This file contains 24321 time-stamped lines. The time stamps range from 67.101133 million years ago (Ma) to 0.000564 Ma. In this paper, we ignore uncertainty in the time stamps. Accounting for the uncertainty in the dating of the observations is beyond the scope of the present study and is left for future research. We order the observations such that the series start at -67.101133 million years and end at -0.000564 million years. Of these 24321 observations, only 23722 fall on distinct time stamps; the difference of 599 arises because some time stamps carry more than one observation (up to four; see below), on either or both series.

\begin{figure}
\centering
\includegraphics[width=.45\textwidth]{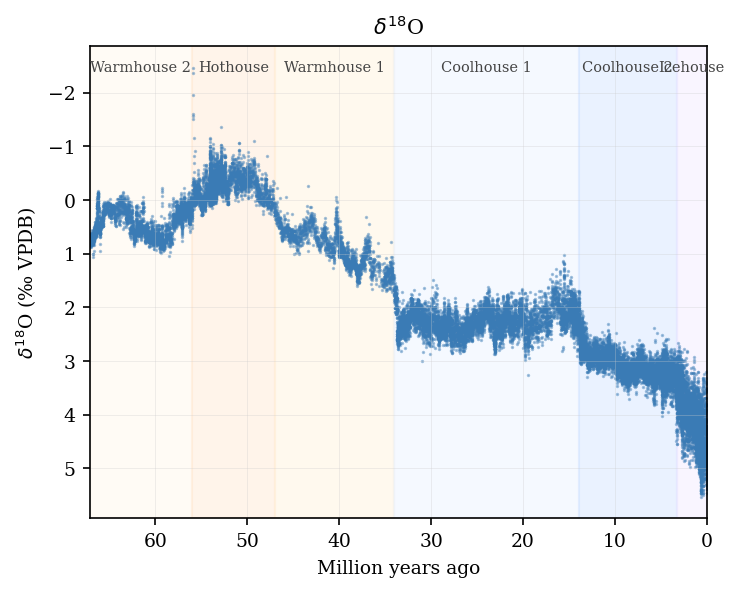}
\includegraphics[width=.45\textwidth]{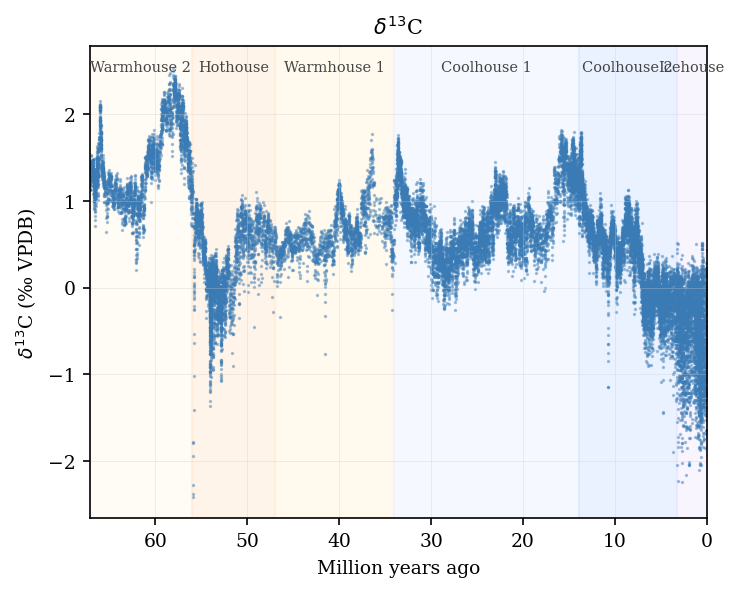}\\
\includegraphics[width=.45\textwidth]{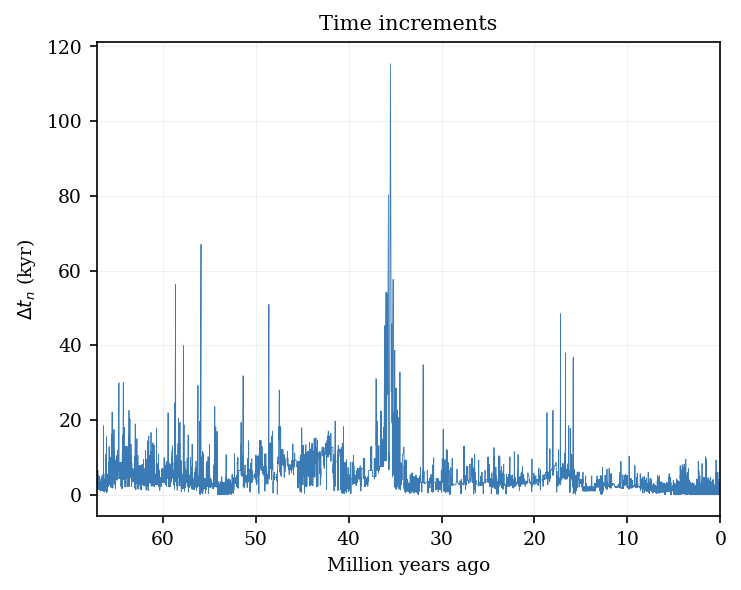}
\caption{\footnotesize Top left panel: $\delta^{18}$O data, top right panel: $\delta^{13}$C data, bottom panel: Time series of differences in the time stamp. The data are from \cite{westerhold2020}. The shaded bands in the top two panels mark the six climate states of \cite{westerhold2020} (Table~\ref{T:Climate_States}). Note that the $y$-axis for $\delta^{18}$O is reversed, following common practice. \label{F:Data}}
\end{figure}

We use $\delta^{18}$O and $\delta^{13}$C observations (column L ``benthic d18O VPDB Corr'' and column K ``benthic d13C VPDB Corr'') from the data file. There are four time stamps for which neither $\delta^{18}$O nor $\delta^{13}$C are available (data file observation numbers 2823, 4895, 4898, 4909). The maximum number of observations at any time stamp is four, often, but not always, originating from different studies.

Figure \ref{F:Data} shows the time series of $\delta^{18}$O (top left panel) and $\delta^{13}$C (top right panel). The bottom panel shows the time series of differences in the time stamps of the observations in million years. The smallest difference that occurs is a single year (data file obs. no. 1778), the biggest is 115366 years (data file obs. no. 16932). The plot in the bottom panel of Figure \ref{F:Data} shows that the data set is denser in the later part of the record and sparser in the earlier.

We show that a data set as complex as this can be analyzed by means of state space methods tailored to the continuous-time nature of the time stamps. Specifically, we employ so-called unobserved component models, and we specify the time series under study as consisting of a set of latent components (signal) plus measurement noise \citep{harvey2000signal,harvey2009unobserved}. Multiple measurements at the same time can be treated. The Kalman filtering and smoothing recursions allow for the estimation of the unobserved components (signal extraction). 

We consider several alternative models that differ in the integration order of the unobserved component. In the simplest case, the random-walk-plus-noise model, the unobserved component is a random walk, that is, integrated of order one. This stochastic trend is a flexible, non-stationary time series model that can accommodate the changing trend directions in the series. With increasing orders of integration of the latent component, the degree of smoothing increases. We describe the relation of the integration orders to orders of a Butterworth filter applied in the frequency domain, and we describe the gain functions implied by the maximum likelihood estimation of our time-domain models \citep{gomez1999,gomez2001,harvey2003}. 

The separate specification of measurement and transition equations in a state space model, with their own respective error processes, allows for reflecting the different sources of the data by way of specifying different measurement equation variances. In the main analysis we differentiate these variances by the ten deep-sea drill sites from which the data originate, and we let each site have its own constant level offset, a bias correction relative to a reference site; alternative specifications that differentiate by the 34 studies of origin or by the 23 benthic foraminifera species are reported in an appendix. The variances of the random variables that drive the transition equation are differentiated according to the six different climate states identified in \cite{westerhold2020}. We consider bivariate models for $\delta^{18}$O and $\delta^{13}$C jointly, such that we can exploit their covariation to inform the estimation of our models.

The rest of the paper is organized as follows: In Section \ref{S:RWN}, we apply random-walk-plus-noise models to the data set, both without and with differentiating variances. In Section \ref{S:Butterworth}, we describe the relation of unobserved component models with the Butterworth filter. In Section \ref{S:IWN}, we explore higher integration orders of the unobserved component, which correspond to higher orders of Butterworth filters. In Section \ref{S:RWNBiv}, we generalize the univariate analysis for $\delta^{18}$O and $\delta^{13}$C to a joint bivariate analysis for both variables, based on the random-walk-plus-noise model.
Section \ref{S:Concl} concludes.

\section{Random walk plus noise models}\label{S:RWN}

We propose the class of unobserved component models for $\delta^{18}$O and $\delta^{13}$C, since these models can accommodate non-stationarity, unevenly spaced observations, missing data, and multiple simultaneous observations for the same object of interest. Unobserved component time series models can be written in univariate form as 
\begin{align} \label{eq:decomp}
y_t &= \mu_t + \psi_t,
\end{align}
where $y_t$ is the observable variable of interest ($\delta^{18}$O or $\delta^{13}$C) at continuous time $t$, $\mu_t$ is the ``unobserved component'' of interest, and $\psi_t$ is a stationary random error. We can think of $\mu_t$ as the ``signal'' and $\psi_t$ as the ``noise.'' Data observations on $y_t$ are at a discrete sequence of time stamps $t_1, t_2, \ldots, t_N$, and we write $y_n=y_{t_n}$. The sequence $\{t_n\}_{n=1,\ldots,N}$ is ordered but irregularly spaced, and there may be repeated elements, i.e., several observations with the same time stamp. The first three and the last three observations in the $\delta^{18}$O and $\delta^{13}$C series are shown in Table \ref{T:Obs} for illustration. The time increments are reported in the third column.

The simplest specification in this class that we consider for the two time series studied here is the random-walk-plus-noise model
\begin{eqnarray}
y_n = \mu_{t_n} + \eps_n, & & \eps_n\stackrel{i.i.d.}{\sim} \mathsf{N}(0,\sigma_\eps^2),\notag\\
\mu_{t_n+\Delta t_n} = \mu_{t_n} + \eta_{t_n}, & & \eta_{t_n}\stackrel{i.i.d.}{\sim} \mathsf{N}(0,\sigma_\eta^2\Delta t_n), \label{E:RWN}
\end{eqnarray}
where the ``signal'' $\mu_{t_n}$ is specified as a random walk process with time increment $\Delta t_n = t_{n+1} - t_n\ge 0$, $n=0,\ldots,N-1$.  The random walk component $\mu_{t_n}$ is non-stationary. If the $(n{+}1)$-th and the $n$-th observation have the same time stamp, it holds that $\Delta t_n = 0$. The noise terms $\eps_n$ and $\eta_{t_n}$ are identically and independently distributed ($i.i.d.$) normal random variables with mean zero. They are assumed to be independent of each other. The random error $\eps_n$ has taken the role of the noise term $\psi$ in the basic signal-plus-noise decomposition in Equation \eqref{eq:decomp} and has variance $\sigma_\eps^2 >0$.  

\begin{table}[h]
\caption{\footnotesize First three and last three observations of the $\delta^{18}$O and $\delta^{13}$C time series with time stamps and time increments. The data are from the accompanying data file of \cite{westerhold2020}, file name \texttt{aba6853\_tables\_s8\_s34.xlsx}, tab ``Table S33'', column L ``benthic d18O VPDB Corr'' and column K ``benthic d13C VPDB Corr''. The data are sorted according to the time stamp in column D ``age\_tuned''. The column ``age in Ma (data file)'' reproduces the positive ages from the data file; the column ``$t_n$ in Ma (model)'' gives the same time stamps in the negative-age convention used throughout the paper, in which the series runs forward from $-67.101133$ to $-0.000564$ Ma. The increment $\Delta t_n = t_{n+1}-t_n$ is the gap to the following observation and is therefore not defined for the final observation. Here $n$ is the unique-time-stamp index used throughout the paper (running to $N=23722$); the column ``entry in data file'' gives the raw line number in the source file, so the two differ by the 599 repeated time stamps.
\label{T:Obs}}
\centering
\begin{tabular}{ccccccc}\hline
$n$     & entry in & age in Ma & $t_n$ in Ma & $\Delta t_n$ & $\delta^{18}$O & $\delta^{13}$C \\
        & data file    & (data file) & (model) &              & \multicolumn{2}{c}{$y_n$} \\
\hline
1       & 24321 & 67.101133 & $-67.101133$ & 0.002158 & 0.800    & 1.376 \\
2       & 24320 & 67.098975 & $-67.098975$ & 0.002157 & 0.800 & 1.343 \\
3       & 24319 & 67.096818 & $-67.096818$ & 0.002158 & 0.800 & 1.320 \\
\vdots  &   & \vdots & \vdots & \vdots & \vdots & \vdots \\
23720   & 3 & 0.003099 & $-0.003099$ & 0.000410 & 3.56 & -0.07 \\
23721   & 2 & 0.002689 & $-0.002689$ & 0.002125 & 3.57 & 0.43 \\
23722   & 1 & 0.000564 & $-0.000564$ & NA & 3.52 & -0.80 \\\hline
\end{tabular}
\end{table}

The variance of the noise term $\eta_{t_n}$ in model \eqref{E:RWN} contains the difference in the time stamps $\Delta t_n$. Loosely speaking, the random variable $\eta_{t_n}$ captures the variation occurring in the process $\mu_{t_n}$ in the time interval $[t_n, t_n+\Delta t_n]$, and so if $\mu_{t_n}$ is a random walk, the variance of this increment is $\sigma_\eta^2\Delta t_n$. The time scaling reflects the intuition that the further apart two observations are, i.e., the larger $\Delta t_n$, the more room there is for the signal to move.

The fixed parameters $\sigma_\eps^2$ and $\sigma_\eta^2$ in model \eqref{E:RWN} are estimated by the method of maximum likelihood, which requires the numerical maximisation of the likelihood function. Given the linear and Gaussian structure of the random-walk-plus-noise model, it follows that the evaluation of the likelihood can be based on a prediction error decomposition and computed in the Kalman filter recursions, see \cite{Durbin2012}, in particular Ch. 7, for a textbook treatment. The first equation in model \eqref{E:RWN} constitutes a measurement equation and the second a transition equation of a state space model. The state variable $\mu_{t}$ is unobserved and inferred from the observations $y_1,\ldots,y_n$. In particular, estimates of the unobserved state $\mu_t$ can be obtained in predicted form ($\E[\mu_{t_n+\Delta t_n} | \{y_s,\; s\le t_n\}]$), filtered form ($\E[\mu_{t_n+\Delta t_n} | \{y_s,\; s\le t_{n+1}\}]$), and smoothed form ($\E[\mu_{t+\Delta t_n} | \{y_s,\; s\le t_N\}]$) from the Kalman filtering (forwards) and smoothing (backwards) recursions, which can handle irregularly spaced and missing observations. We implement parameter estimation and state prediction, filtering, and smoothing in \texttt{Python} and \texttt{Fortran} and in \texttt{OxMetrics/SsfPack} \citep{doornik2021,koopman2008ssfpack}. For all these tasks, we treat the initial state $\mu _0$ as unknown and employ a diffuse initialisation.

\begin{figure}[htbp]
\centering
\includegraphics[width=.45\textwidth]{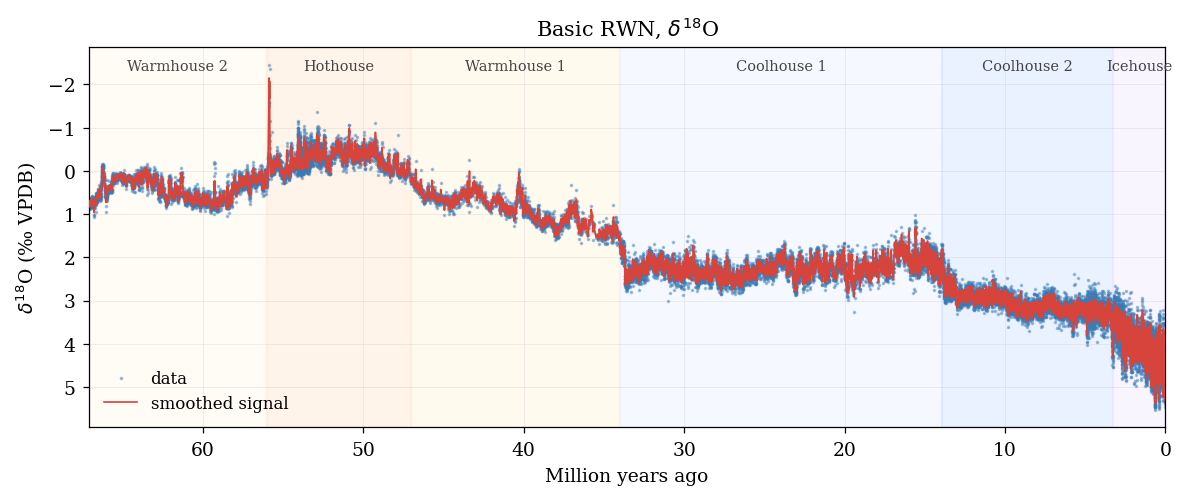}
\includegraphics[width=.45\textwidth]{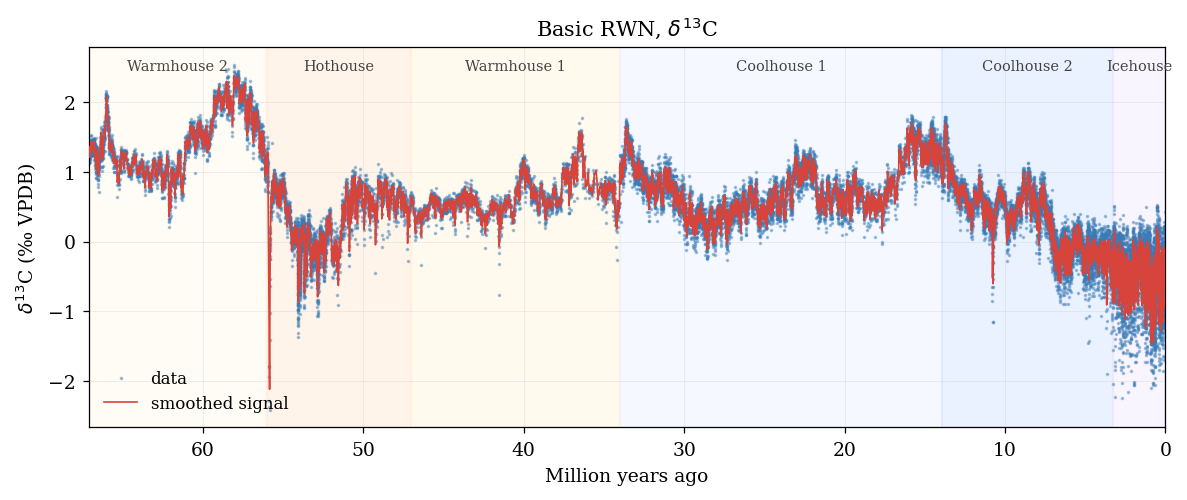}\\
\includegraphics[width=.45\textwidth]{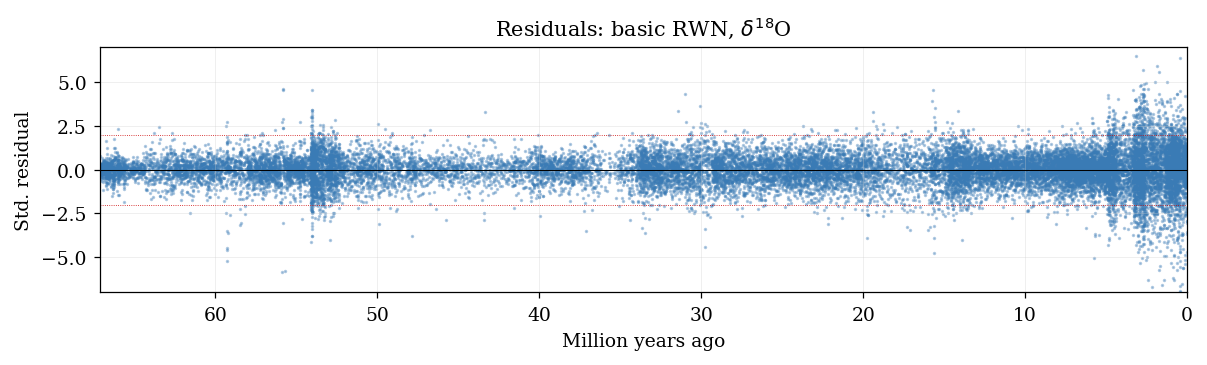}
\includegraphics[width=.45\textwidth]{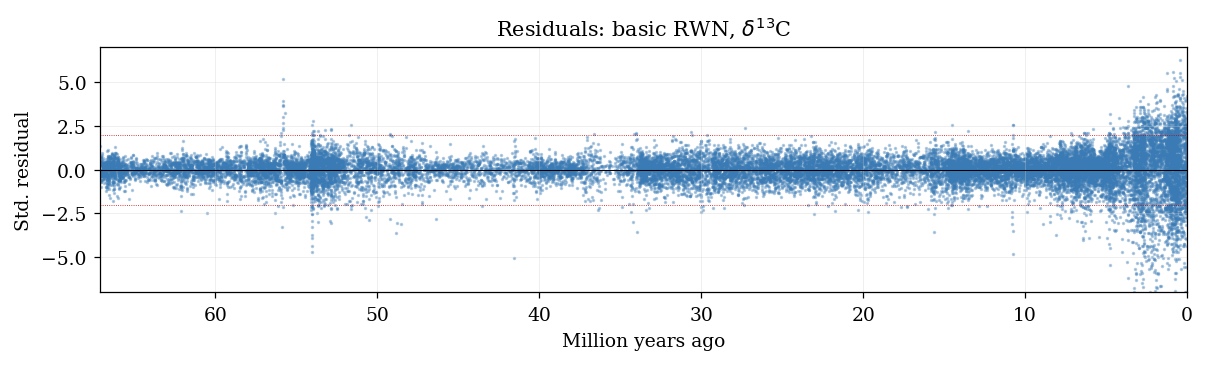}\\[2mm]
\includegraphics[width=.45\textwidth]{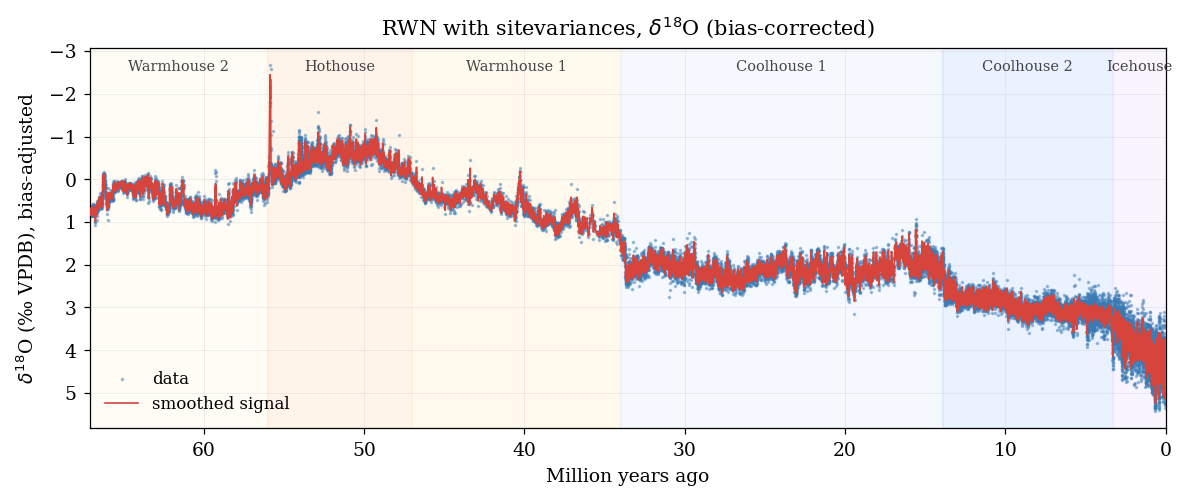}
\includegraphics[width=.45\textwidth]{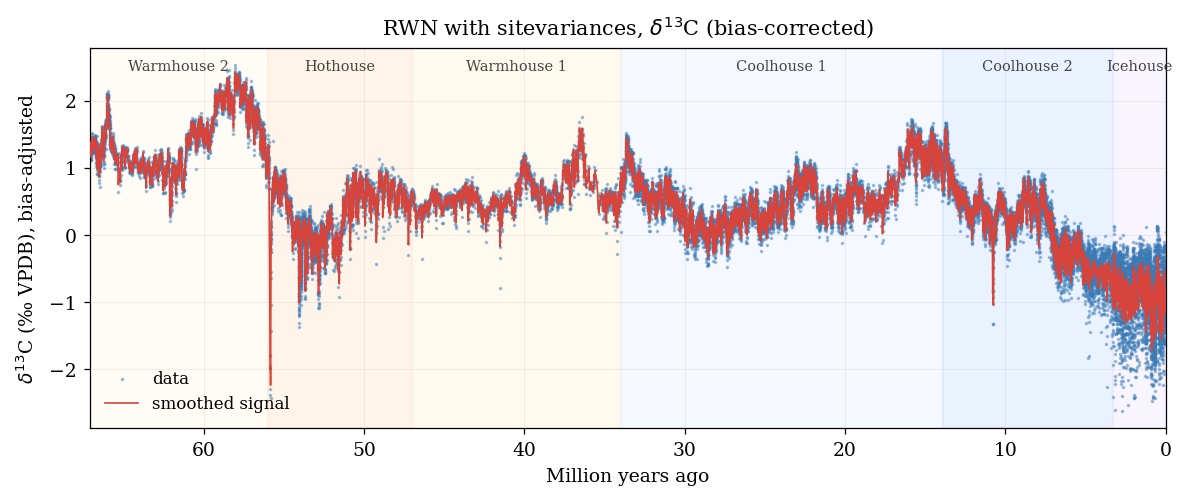}\\
\includegraphics[width=.45\textwidth]{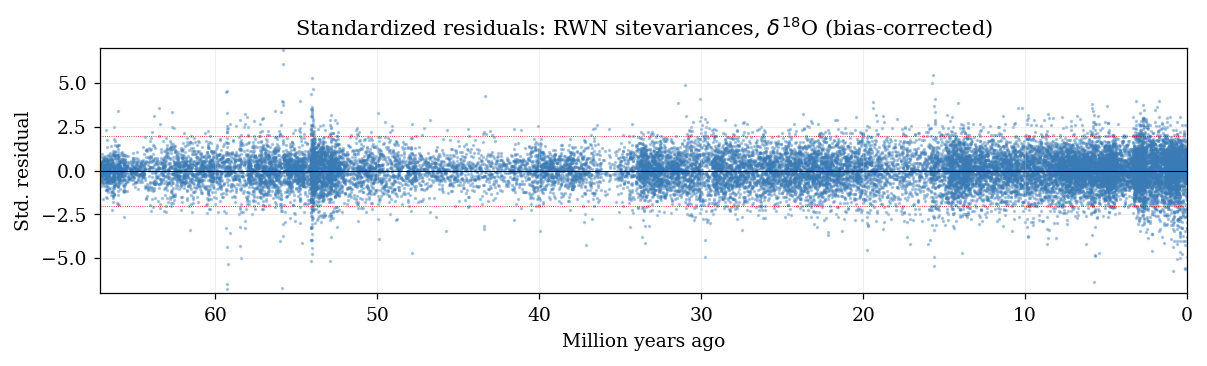}
\includegraphics[width=.45\textwidth]{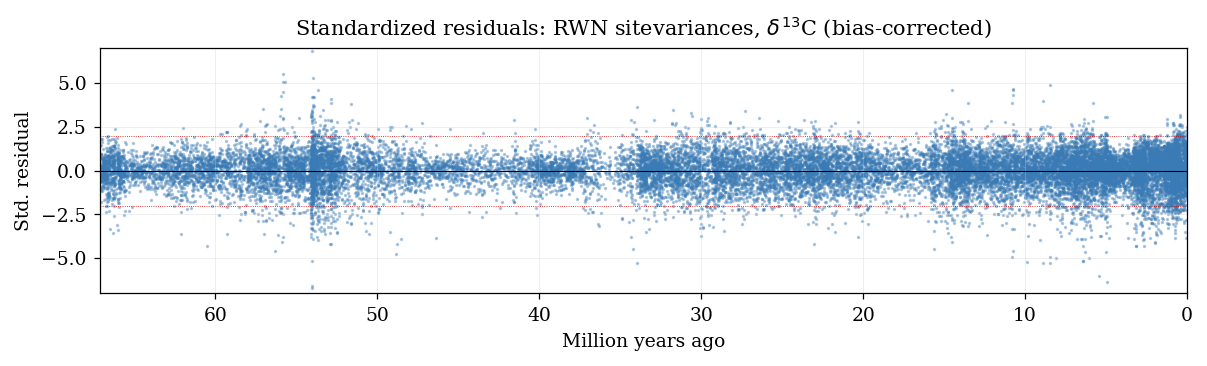}\\
\caption{\footnotesize Rows 1--2: random-walk-plus-noise model \eqref{E:RWN}. Top row: $\delta^{18}$O data (blue) and smoothed state $\E[\mu_{t+\Delta t} | \{y_s,\; s\le t_N\}]$ (red) from Kalman smoothing recursions (left), and the same for $\delta^{13}$C (right). Second row: standardized prediction residuals for $\delta^{18}$O (left) and $\delta^{13}$C (right). Rows 3--4: the same arrangement for model \eqref{E:RWN_SitV} with site-specific constant intercepts $c_s$ and measurement variances $\sigma_{\eps,s}^2$ (proxy panels on top, residuals beneath). Note that the $y$-axis for $\delta^{18}$O is reversed, following common practice.\label{F:RWN}}
\label{F:RWN_SitV}
\end{figure}

The two graphs in the top row of Figure \ref{F:RWN} show the data for $\delta^{18}$O and $\delta^{13}$C together with their smoothed estimates of $\mu_{t_n}$ obtained from the Kalman filter and smoothing recursions applied to model \eqref{E:RWN}, with $\sigma _\eps ^2$ and $\sigma _\eta ^2$ set to their respective maximum likelihood estimated values. The two graphs in the second row show the standardized prediction residuals as computed using the Kalman filter output \citep[p. 38, eq. (2.65)]{Durbin2012}. Clearly, the residuals depicted in Figure \ref{F:RWN} do not appear as an $i.i.d.$ sequence with constant variance. Therefore, while the standard errors obtained from the estimation process give some sense of the dispersion of the parameter estimates, they cannot be interpreted in the sense of a central limit theorem. With this caveat, the maximum likelihood estimates of model \eqref{E:RWN} are reported in Table \ref{T:RWN}. The Bayes information criterion
\[
\text{BIC} = P \log N - 2\log L,
\]
where $P$ is the number of parameters in the model, $N$ is the number of observations, and $L$ is the maximized value of the likelihood function, is a measure of fit that is penalized for the number of estimated parameters. Models with small values of the BIC are therefore preferred. In this first model, the number of estimated parameters is only two. However, the penalty term becomes more important when we increase the complexity of the model, as we do in the following.

\begin{table}[h]
\caption{\footnotesize Maximum likelihood parameter estimates for model \eqref{E:RWN}. Standard errors in parentheses.
Maximised log-likelihood values are reported and the Bayes information criterion (BIC).\label{T:RWN}}
\centering
\begin{tabular}{ccc}\hline
 & $\delta^{18}$O & $\delta^{13}$C \\\hline
$\sigma_\eps^2$ & 0.0205 (0.0002) & 0.0340 (0.0004) \\
$\sigma_\eta^2$ & 1.8364 (0.0480) & 1.2135 (0.0403) \\\hline
log-likelihood & 7218.57 & 2992.10 \\
BIC             & -14416.99 & -5964.05 \\\hline
\end{tabular}
\end{table}

\subsection{The measurement variance as a function of drill sites}\label{S:RWN_Sites}

The data set prepared in \cite{westerhold2020} originates from ten drill sites in the deep ocean (column ``Record'' in tab ``Table S33'' of the file \linebreak \texttt{aba6853\_tables\_s8\_s34.xlsx}): 1146, 1218, 1258, 1262, 1263, 1264, 1265, U1337, U1338, and the CearaRiseStack composite. Their geographical locations are shown in Figure~S1 of the supplementary material of \cite{westerhold2020}. Because the measurement error reflects, in part, properties of the core itself (preservation, sampling resolution, local oceanographic noise), it is natural to allow the measurement variance to depend on the drill site from which the observation was taken. We also allow each drill site to have its own constant level offset $c_s$ relative to a reference site, capturing systematic calibration differences between cores (e.g., due to bottom-water properties or coring depth that affect the level but not the dynamics of the records). The measurement equation in model \eqref{E:RWN} therefore becomes
\begin{equation}
y_{n} = \mu_{t_n} + c_s + \eps_{n}, \qquad \eps_{n}\stackrel{indep.}{\sim} \mathsf{N}(0,\sigma_{\eps,s}^2),
\qquad s= s(n) \in \{1,\ldots,10\},\label{E:RWN_SitV}
\end{equation}
where $s$ identifies the drill site of observation $y_n$. Because the underlying state $\mu_{t_n}$ is a non-stationary random walk, only differences in $c_s$ are identifiable; we therefore normalize $c_{s_0} = 0$ for a reference site $s_0$ (Site 1262, with the most observations in the early record). The remaining $c_s$ enter the measurement equation as level offsets that are constant in time, and are estimated by maximum likelihood jointly with the variance parameters; equivalently, they may be treated as diffuse constant states and recovered within the Kalman filtering and smoothing recursions, which gives identical estimates. The parameter vector therefore has twenty elements: nine biases $c_s$ for $s \neq s_0$, ten $\sigma_{\eps,s}^2$, and one $\sigma_{\eta}^2$.



The lower half of Figure~\ref{F:RWN_SitV} (rows 3--4) shows the data and the smoothed values of $\mu_t$ from the Kalman filter and smoothing recursions applied to model \eqref{E:RWN_SitV}, evaluated at the maximum likelihood estimates. Data are shown after subtracting the estimated site-specific intercepts $\hat c_s$, so that all sites are centered on the common signal. The standardized prediction residuals (bottom row) display visibly less time variation in their variance than under the constant-variance model \eqref{E:RWN} in the upper half of the same figure. Table~\ref{T:RWN_SitV} in the appendix reports the estimates, together with the maximized log-likelihood and BIC for each proxy. The improvement in fit over the constant-variance model \eqref{E:RWN} is substantial: the BIC falls from $-14417$ to $-19516$ for $\delta^{18}$O and from $-5964$ to $-18021$ for $\delta^{13}$C. Most sites have $\sigma_{\eps,s}^2$ between $0.006$ and $0.018$ for both proxies; the ``CearaRiseStack'' composite, which aggregates a number of older studies, has by far the largest noise variance (about $0.055$ for $\delta^{18}$O and $0.129$ for $\delta^{13}$C). The site intercepts $\hat c_s$ are tightly identified for sites that share calendar time with the reference site 1262 (e.g., site 1263 with $\hat c_s = +0.21\permil$ in $\delta^{18}$O, SE $0.006$); sites that do not overlap with 1262 still have identifiable biases through transitive overlap with intermediate sites, but with larger standard errors. The biases jointly improve the log-likelihood by about 530 points for $\delta^{18}$O (an LR statistic of $\sim 1060$ against $\chi^2_9$) and by about 35 points for $\delta^{13}$C (LR $\sim 71$), confirming that systematic level offsets between cores are a significant feature of the data, especially for $\delta^{18}$O.

Appendix~\ref{A:Sp} reports results on alternative specifications where measurement equation variances are differentiated according to the studies they were published in and according to the foraminifera species.

\subsection{The transition variance as a function of climate states}

\begin{table}
\caption{\footnotesize Climate states identified in \cite{westerhold2020}. The column ``unique obs'' gives the range of the unique-observation index $n$ (of the $N=23722$ unique time stamps) falling in each climate state; ``Obs (data file)'' gives the corresponding range of line numbers in the original data file. \label{T:Climate_States}}
\centering
\footnotesize
\begin{tabular}{lclrr}\hline
State & Million years ago & Climate state & unique obs & Obs (data file) \\\hline
1 & 67.10113 to 56  & Warmhouse 2   & 1--2690       & 21561--24321 \\
2 & 56 to 47        & Hothouse      & 2691--5693    & 18531--21560 \\
3 & 47 to 34        & Warmhouse 1   & 5694--7471    & 16744--18530 \\
4 & 34 to 13.9      & Coolhouse 1   & 7472--13933   & 10067--16743 \\
5 & 13.9 to 3.3     & Coolhouse 2   & 13934--20179  & 3762--10066 \\
6 & 3.3 to 0.000564 & Icehouse      & 20180--23722  & 1--3761 \\\hline
\end{tabular}
\end{table}

\cite{westerhold2020} identify different climate states in the records of $\delta^{18}$O and $\delta^{13}$C with the help of recurrence analysis. The corresponding periods are shown in Table \ref{T:Climate_States}. For example, the data plots show that the variance of the observations seem to increase towards the later part of the sample. This is particularly conspicuous for the ``icehouse'' regime in the $\delta^{13}$C record.

We employ these periods and differentiate the variances driving the unobserved component $\mu_t$ according to climate states. We place this differentiation in the transition equation rather than in the measurement equation: there are physical reasons to expect that the signal $\mu_t$ itself evolves at different speeds in different climate states, and testing this hypothesis requires the transition-equation variance to depend on the climate state. This also accords with our aim for $\mu_t$ to carry a physical interpretation as a climate-state variable that is observed in $y_t$ with measurement noise.

The transition equation in model \eqref{E:RWN} then becomes
\begin{equation}
\mu_{t_n+\Delta t_n} = \mu_{t_n} + \eta_{t_n},\qquad \eta_{t_n}\stackrel{i.i.d.}{\sim} \mathsf{N}(0,\sigma_{\eta,k}^2\Delta {t_n}),\qquad
k= k(n) \in \{1,\ldots, 6\}, \label{E:RWN_SPV}
\end{equation}
where index $k$ refers to one of the six climate state regimes presented in Table \ref{T:Climate_States}.
For example, if $n\in \{1,\ldots,2690\}$, we have $k=1$,
if $n\in \{2691,\ldots, 5693\}$, we have $k=2$, and so forth.

In the measurement equation, we differentiate the variances in two alternative ways. First, we employ the drill-site specification \eqref{E:RWN_SitV} from Section~\ref{S:RWN_Sites}, with site-specific variances $\sigma_{\eps,s}^2$ and constant intercepts $c_s$ (with the reference site fixed at $c_{s_0} = 0$). Combined with the six period transition variances $\sigma_{\eta,k}^2$, this specification has 25 parameters and is reported in Table~\ref{T:RWN_SitePV} in the appendix.

Second, we employ the study-based specification (Appendix~\ref{A:Sp}) so that the variances are differentiated according to the studies of origin (the entries in column ``benthic Source'' of the data file). This specification uses 34 study variances. Because each drill site is typically covered by several studies, the studies specification is more granular than the sites specification but uses a larger number of parameters.


The maximum likelihood estimates of the resulting models are reported in Table \ref{T:RWN_SitePV} in the appendix for the drill-site specification \eqref{E:RWN_SitV} with site-specific intercepts, and in Appendix~\ref{A:Sp} for the study specification. Both specifications produce climate-state dependent transition variances that vary strongly across periods; by far the highest variance is estimated in the ``icehouse'' period (3.3 to 0.0006 Ma), corresponding to the visual impression of the late record mentioned above. The site-based specification captures most of the heterogeneity across observations with 10 measurement variances plus 9 site intercepts; the most pronounced site is the ``CearaRiseStack'' composite (in particular for $\delta^{13}$C), which aggregates several older studies. Adding the climate-state period structure to the site model increases the log-likelihood by about 840 points (from $9858.93$ in Table~\ref{T:RWN_SitV} to $10701.27$) for $\delta^{18}$O and by about 470 points for $\delta^{13}$C, with the lowest BIC in the univariate RWN class. The site intercepts $\hat c_s$ in Table~\ref{T:RWN_SitePV} are very close to those obtained without period differentiation in Table~\ref{T:RWN_SitV} in the appendix, indicating that the static calibration offsets are orthogonal to the regime structure in the transition equation. The BIC is lower (better) for the study specification, reflecting the larger amount of granular information in 34 measurement variances; however, the site specification provides a more parsimonious description directly tied to the geographical location of the deep-sea cores.

\section{Butterworth filtering: time domain interpretation}\label{S:Butterworth}

The models proposed in the previous section naturally lend themselves to an interpretation where a signal $\mu_t$ is extracted from observations $y_t$ and distinguished from noise $\psi_t$. In this section, we present the well-known connection to signal extraction in the frequency domain and show how the random-walk-plus-noise models of the previous section and the integrated-random-walk-plus-noise models of the next section are time-domain counterparts of the Butterworth filter. For ease of notation, we use $t$ for discrete time in this section.

Consider a time series that consists of an autoregressive component $\mu_t = \phi_1 \mu_{t-1} + \ldots + \phi_p \mu_{t-p} + \eta_t$, where $\eta_t$ is i.i.d. with mean zero and variance $\sigma_\eta^2$, plus an i.i.d. contamination $\eps_t$ with mean zero and variance $\sigma_\eps^2$ that is independent of $\eta_t$:
\begin{equation}\label{E:UC1}
y_t = \Phi^{-1}(L)\eta_t + \eps_t,
\end{equation}
where $\Phi(L)=1-\phi_1 L -\ldots - \phi_p L^p$ is the lag polynomial corresponding to the autoregressive component with roots outside the unit circle. For a zero-mean stationary time series such as this, define the autocovariance-generating function
\[
g_y(z) = \sum_{j=-\infty}^\infty \gamma(j) z^j,\; z\in\C,
\]
with $\gamma(j)=\E(y_t y_{t-j})$. Since $\eta_t$ and $\eps_t$ are independent, we have that
\[
g_y(z) = g_\mu(z) + g_\eps(z),
\]
where
\[
g_\mu(z) = \frac{\sigma_\eta^2}{|\Phi(z)|^2} = \frac{\sigma_\eta^2}{\Phi(z)\Phi(z^{-1})},
\]
and $g_\eps(z)=\sigma_\eps^2$. The convention $|\Phi(z)|^2=\Phi(z)\Phi(z^{-1})$ is motivated by the fact that we consider $z\in\C$ only on the unit circle, i.e., $z=\exp(-i\lambda)$, $\lambda\in\R$, and all $\gamma(j)\in\R$, thus $|g(z)|^2=g(z)\overline{g(z)}=g(z)g(\bar{z})$, cf. \cite{Whittle1983}, p. 12.

The Wiener-Kolmogorov filter for the doubly-infinite sample for model \eqref{E:UC1} is given by
\[
\hat\mu_t = \sum_{j=-\infty}^\infty b_j y_{t-j} = B(L) y_t,
\]
with response function
\[
B(z) = \frac{g_\mu(z)}{g_\mu(z) + g_\eps(z)}  = \frac{1}{1+q^{-1}|\Phi(z)|^2},
\]
where $q=\sigma_\eta^2/\sigma_\eps^2$ is the signal-to-noise ratio. For the integrated models considered below we index the response function, the gain function, and the signal-to-noise ratio by the integration order $m$, writing $B_m(z)$, $G_m(\lambda)$, and $q_m$, so that higher $m$ (a smoother extracted signal) is made explicit in the notation. For the AR(1) case $\Phi(z) = 1-\phi z$, and $\phi\to 1$, we get the random-walk-plus-noise model
\[
y_t = \sum_{j\le t} \eta_j + \eps_t = (1-L)^{-1}\eta_t + \eps_t.
\]
\cite{bell1984} showed that the Wiener-Kolmogorov filter is asymptotically valid in the nonstationary case under standard assumptions, see also \cite{gomez1999}. The random-walk-plus-noise model therefore has response function
\[
B_1(z) = \frac{1}{1+q_1^{-1}|1-z|^2},
\]
where the subscript records the integration order $m=1$ and $q_1\equiv q=\sigma_\eta^2/\sigma_\eps^2$ is the signal-to-noise ratio defined above. For a general integration order $m$ the signal-to-noise ratio is $q_m=\sigma_{\eta,m}^2/\sigma_\eps^2$; as explained below, its maximum-likelihood estimate depends on $m$. The integrated random-walk-plus-noise model,
\[
y_t = \sum_{s\le t} \sum_{j\le s} \eta_j + \eps_t = (1-L)^{-2}\eta_t + \eps_t,
\]
has response function
\[
B_2(z) = \frac{1}{1+q_2^{-1}|1-z|^4}.
\]
Integrating the random walk $m$ times yields
\[
y_t = (1-L)^{-m}\eta_t + \eps_t,
\]
with response function
\[
B_m(z) = \frac{1}{1+q_m^{-1}|1-z|^{2m}}.
\]
For $z=e^{-i\lambda}$, we then obtain the gain function $G_m(\lambda)$ of the $m$-th order integrated random-walk-plus-noise model
\[
G_m(\lambda) = |B_m(e^{-i\lambda})| = \frac{1}{1+q_m^{-1}[(1-e^{-i\lambda})(1-e^{i\lambda})]^m}.
\]
Since the filter is symmetric, and the response function is non-negative, the gain function is equal to the response function. With $(1-e^{-i\lambda})(1-e^{i\lambda})=4\sin^2(\lambda/2)$, we have that
\begin{equation}
G_m(\lambda) = \frac{1}{1+q_m^{-1} 2^{2m}\sin^{2m}(\lambda/2)},\label{E:gain}
\end{equation}
and, if $q_m \le 2^{2m}$, letting
\begin{equation}
\lambda_h := 2\sin^{-1}\left(q_m^{\frac{1}{2m}}/2\right),\label{E:cutoff}
\end{equation}
we can write
\[
G_m(\lambda) = \frac{1}{1+\left(\frac{\sin \frac{\lambda}{2}}{\sin \frac{\lambda_h}{2}}\right)^{2m}}.
\]
Here the subscript $h$ stands for ``half power'': $\lambda_h$ is the frequency at which the gain has fallen to one half, $G_m(\lambda_h)=\tfrac12$, as is seen by setting $\lambda=\lambda_h$ in the display above. It is the cutoff frequency of the filter and is unrelated to the integration order $m$. This is the gain function of the Butterworth filter of order $m$ with cutoff frequency $\lambda_h/2$, see \cite{harvey2000signal,harvey2003}. Estimating the $m$-th order integrated random-walk-plus-noise model by maximum likelihood and obtaining variance estimates $\hat\sigma_\eta^2$ and $\hat\sigma_\eps^2$ thus implies an estimated cutoff frequency
\[
\frac{\hat\lambda_h}{2} = \sin^{-1}\left(\hat q_m^{\frac{1}{2m}}/2\right).
\]
The higher the order $m$ of the filter, the smoother the filtered signal $\hat\mu_t$, the smaller the estimated variance $\hat\sigma_\eta^2$ relative to $\hat\sigma_\eps^2$ and the lower the estimated signal-to-noise ratio $\hat q_m$. In this sense, $q_m$ depends on $m$.

\section{Integrated random walk plus noise models}\label{S:IWN}

The insights of the previous section into the connection of integrated random walk models ($m>1$) with the Butterworth filter suggest considering models $y_n = \mu_{t_n} + \eps_n$, with $\eps _n\stackrel{i.i.d.}{\sim}\mathsf{N}(0,\sigma_\eps^2)$, where the latent level $\mu_t$ is an $m$-fold integrated Wiener process \citep{weckerAnsley1983}. These models smooth the data to a higher degree than the random-walk-plus-noise model ($m=1$), since the filter weighs lower frequencies more with increasing $m$.

In continuous time, the latent level $\mu_t$ and its first $m-1$ time derivatives form a state vector of dimension~$m$. Let $\mu_t^{(j)} = d^j\mu_t/dt^j$ denote the $j$-th time derivative, with $\mu_t^{(0)} = \mu_t$. The continuous-time dynamics are
\[
d\mu_t^{(j)} = \mu_t^{(j+1)}\, dt, \quad j=0,\ldots,m-2, \qquad d\mu_t^{(m-1)} = \sigma_{\eta,m}\, dW_t,
\]
where $W_t$ is a standard Wiener process. That is, the $(m-1)$-th derivative is a Brownian motion, and each lower-order derivative is obtained by integration. We refer to the resulting latent process as \emph{integrated white noise of order $m$}, IWN($m$).

\paragraph{IWN(2).} For $m=2$, the state vector is $(\mu_t,\, \mu_t^{(1)})'$, consisting of the level and its first derivative (the slope). The exact discrete-time transition over an interval $\Delta t_n = t_{n+1}-t_n$ is
\begin{equation}
\left[\begin{array}{c}
\mu_{t_n+\Delta t_n} \\ \mu_{t_n+\Delta t_n}^{(1)}
\end{array}\right]
=
\left[\begin{array}{cc}
1 & \Delta t_n \\ 0 & 1
\end{array}\right]
\left[\begin{array}{c}
\mu_{t_n} \\ \mu_{t_n}^{(1)}
\end{array}\right]
+
\left[\begin{array}{c}
\eta_{1,t_n} \\ \eta_{2,t_n}
\end{array}\right],\label{E:IWN}
\end{equation}
where the state disturbance vector has covariance
\[
\text{Var}\!\left[\begin{array}{c}
\eta_{1,t_n} \\ \eta_{2,t_n}
\end{array}\right]
= \sigma_{\eta,2}^2
\left[\begin{array}{cc}
\frac{\Delta t_n^3}{3} & \frac{\Delta t_n^2}{2} \\[4pt]
\frac{\Delta t_n^2}{2} & \Delta t_n
\end{array}\right].
\]
The off-diagonal terms arise because the level disturbance $\eta_{1,t_n}$ is the integral $\int_{t_n}^{t_{n+1}}\!\int_{t_n}^{s} dW_u\, ds$ of the slope's Brownian increment over $[t_n, t_{n+1}]$.
The measurement equation is as before
\[
y_n = \mu_{t_n} + \eps_n, \quad \eps_n\stackrel{i.i.d.}{\sim}\mathsf{N}(0,\sigma_\eps^2),
\]
and the distinctions according to studies in the measurement equation and according to periods in the transition equation can be introduced as outlined in Section \ref{S:RWN}.

\paragraph{General $m$.} For integration order $m \ge 2$, the state vector is $(\mu_t,\, \mu_t^{(1)},\, \ldots,\, \mu_t^{(m-1)})'$. The exact discrete-time transition is
\begin{equation}
\left[\begin{array}{c}
\mu_{t_n+\Delta t_n} \\ \mu_{t_n+\Delta t_n}^{(1)} \\ \vdots \\ \mu_{t_n+\Delta t_n}^{(m-2)} \\ \mu_{t_n+\Delta t_n}^{(m-1)}
\end{array}\right]
=
T(\Delta t_n)
\left[\begin{array}{c}
\mu_{t_n} \\ \mu_{t_n}^{(1)} \\ \vdots \\ \mu_{t_n}^{(m-2)} \\ \mu_{t_n}^{(m-1)}
\end{array}\right]
+
\boldsymbol{\eta}_{t_n},\qquad \boldsymbol{\eta}_{t_n}\sim \mathsf{N}(\mathbf{0},\, Q(\Delta t_n)),\label{E:IWN_m}
\end{equation}
where the $m\times m$ transition matrix $T(\Delta t_n)$ and covariance matrix $Q(\Delta t_n)$ have entries
\begin{align*}
T(\Delta t_n)_{j+1,k+1} &= \frac{\Delta t_n^{\,k-j}}{(k-j)!}\, \mathbf{1}_{\{k\ge j\}}, \\[2pt]
Q(\Delta t_n)_{j+1,k+1} &= \frac{\sigma_{\eta,m}^2\;\Delta t_n^{\,2m-1-j-k}}{(m\!-\!1\!-\!j)!\,(m\!-\!1\!-\!k)!\,(2m\!-\!1\!-\!j\!-\!k)},
\end{align*}
for $j,k = 0,\ldots, m-1$, so that the row and column indices $j+1$ and $k+1$ run over $1,\ldots,m$, and $\mathbf{1}_{\{\cdot\}}$ denotes the indicator function. The transition matrix is upper triangular with entries given by the Taylor expansion coefficients; the covariance matrix is the variance of the $m$-fold integrated Wiener process increment over $[t_n, t_{n+1}]$, depending on a single parameter $\sigma_{\eta,m}^2$.
We employ these models for the data series $\delta^{13}$C and $\delta^{18}$O next.

\subsection*{Revisiting the analyses of $\delta^{13}$C and $\delta^{18}$O}

We apply this class of models to $\delta^{18}$O and $\delta^{13}$C by differentiating the measurement-equation variances according to drill sites (with the study-based specification relegated to the appendix) and the transition-equation variances according to the six climate states, exactly as for the random-walk-plus-noise model in Section~\ref{S:RWN}.

The order of integration $m$ in model \eqref{E:IWN_m}, equal to the order of the corresponding Butterworth filter, is a design choice by the researcher that reflects the degree of desired smoothing. Given this choice, the estimation maximizes the log-likelihood of model \eqref{E:IWN_m} with respect to the variance parameters by way of the Kalman filter recursions \citep{Durbin2012}; conditional on the data and $m$, the resulting signal-to-noise ratios and the implied gain functions from Equation \eqref{E:gain} are optimal in this sense. The maximized log-likelihood necessarily decreases with $m$, reflecting the higher degree of smoothing of the latent process $\mu_t$: a smoother $\mu_t$ captures less of the data variation, leaving more to the measurement equation. The per-state signal-to-noise ratios, reported as $\hat q_{m,k}=\hat Q_{00}/\bar\sigma_\eps^2$ in Table~\ref{T:IWN_Sites_PV} in the appendix, decline steeply with $m$, which is precisely the increasing smoothing imposed by higher integration orders (the dimension-stable quantity $\hat Q_{00}$ is defined in Equation~\eqref{E:Q00}).



For high integration orders the implied Butterworth cutoff frequency lies very close to zero, so the latent process is strongly smoothed. Nevertheless, the resulting estimated latent processes $\mu$ still show a good fit and a discernible difference in the degree of smoothing for orders $m$ between 4 and 6, as we now demonstrate in the estimation that distinguishes measurement equation variances according to studies and transition equation variances according to climate states.

We employ the same two strategies for differentiating measurement equation variances as in Section~\ref{S:RWN}: by drill site (Equation~\eqref{E:RWN_SitV}, ten variances with site-specific constant intercepts $c_s$) and by study (Appendix~\ref{A:Sp}, 34 variances). For the transition equation, we differentiate variances according to climate states as defined in Table~\ref{T:Climate_States}. We modify Equation \eqref{E:IWN_m} by replacing the single noise rate $\sigma_{\eta,m}^2$ in the covariance matrix $Q(\Delta t_n)$ with climate-state-dependent rates:
\begin{equation}
\sigma_{\eta,m}^2 \;\longrightarrow\; \sigma_{\eta,m,k}^2, \quad k\in\{1,\ldots,6\}, \quad m\in\{2,4,6,8\}.\label{E:IWN_SPV}
\end{equation}
Tables~\ref{T:IWN_Sites_d18O} and~\ref{T:IWN_Sites_d13C} in the appendix report the estimated measurement equation variances, and Table~\ref{T:IWN_Sites_PV} in the appendix the estimated transition equation variances, for the drill-site specification. Figures~\ref{F:IRW_Sites_d18O} and~\ref{F:IRW_Sites_d13C} in the appendix show the estimated latent state processes $\mu_t$ for $\delta^{18}$O and $\delta^{13}$C, respectively, with the corresponding standardized prediction residuals shown beneath each smoothed-state panel. There is no visually discernible difference in the estimated latent processes between orders $m=6$ and $m=8$, and the differences in the estimated measurement equation variances between $m=6$ and $m=8$ are small and within the variation indicated by the estimated standard errors. The estimated transition-equation variance \emph{rates} $\hat\sigma^2_{\eta,m,k}$, by contrast, grow by many orders of magnitude as $m$ increases, reaching $\mathcal{O}(10^{26})$ at $m=8$ (Table~\ref{T:IWN_Sites_PV}). As we explain next, this growth reflects the continuous-time units in which the rate is measured, not the magnitude of the latent process, and the dimension-stable per-step level innovation declines with $m$.

Between $m=4$ and $m=6$, in contrast, there is a visually discernible difference in the estimated latent state processes $\mu_t$ for $\delta^{18}$O and $\delta^{13}$C, and the estimated variances, both in measurement and transition equation, differ substantially. We conclude that integration orders up to $m=6$ are meaningful for $\delta^{18}$O and $\delta^{13}$C. Comparing log-likelihoods across $m$ is not meaningful, since $m$ is a design choice. For given $m$, the site-differentiated specification (Tables~\ref{T:IWN_Sites_d18O} and~\ref{T:IWN_Sites_d13C} in the appendix) attains a substantially higher log-likelihood than a single constant measurement and transition variance, confirming that allowing for site- and climate-state-dependent variances substantially improves the model fit. As for the random-walk-plus-noise model, differentiating the measurement variances by study rather than by site yields a slightly higher log-likelihood at the cost of more parameters, but the site specification provides a more parsimonious description that is directly tied to the geographical locations of the deep-sea cores.

\paragraph{Interpreting the transition-variance estimates.}
The maximum-likelihood transition variances $\hat\sigma^2_{\eta,m,k}$ in Table~\ref{T:IWN_Sites_PV} in the appendix grow by many orders of magnitude as the integration order $m$ increases, reaching $\mathcal{O}(10^{26})$ at $m=8$. This is not a numerical pathology but a direct consequence of the continuous-time parameterization. In the pure $m$-fold IWN the single white-noise innovation enters at the $m$-th derivative, and $\sigma^2_{\eta,m}$ is its variance \emph{rate}, with units $[\text{signal}]^2\,[\text{time}]^{-(2m-1)}$. The variance it injects into the observed \emph{level} over a sampling interval $\Delta t$ is not $\sigma^2_{\eta,m}$ but
\begin{equation}
Q_{00}(\Delta t) \;=\; \sigma^2_{\eta,m}\,\frac{\Delta t^{\,2m-1}}{(2m-1)\,[(m-1)!]^2},
\label{E:Q00}
\end{equation}
because reaching the level from the $m$-th derivative requires $m$ successive integrations, each over the short interval $\Delta t$. With a median spacing $\overline{\Delta t}\approx 0.002$~Ma, the factor $\Delta t^{2m-1}$ is astronomically small ($\Delta t^{15}\approx 10^{-40}$ at $m=8$), so an $\mathcal{O}(10^{-4})$ level innovation requires an $\mathcal{O}(10^{26})$ rate; equivalently, rescaling time so that $\overline{\Delta t}=1$ would render all the values $\mathcal{O}(1)$. The latent process itself does not grow: the smoothed states (Figures~\ref{F:IRW_Sites_d18O} and~\ref{F:IRW_Sites_d13C} in the appendix) remain $\mathcal{O}(1)$ and are pinned to the data. The dimension-stable companions reported alongside the rate (the per-step level-innovation standard deviation $\sqrt{\hat Q_{00}}$ and the signal-to-noise ratio $\hat q_{m,k}=\hat Q_{00}/\bar\sigma^2_\eps$) decline steeply with $m$, and that decline is precisely the increasing smoothing imposed by higher integration orders.






\section{Bivariate models}  \label{S:RWNBiv}

The covariation in $\delta^{18}$O and $\delta^{13}$C allows for an improvement of the estimation of the unobserved $\mu$-components in $\delta^{18}$O and $\delta^{13}$C by modelling the two time series jointly and estimating the covariance of the error terms $\eta$.

To conceptualize the time dependence in the covariances of increments of  $\mu^{\delta^{18}O}_{t}$ and $\mu^{\delta^{13}C}_{t}$, we consider a stylized situation for Brownian motion in continuous time.  Let $B_1(t)$ and $B_3(t)$ be two independent Brownian motions. Let $\rho\in (-1,1)$ be a correlation parameter, and let
\[
B_2(t) = \rho B_1(t) + \sqrt{1-\rho^2} B_3(t).
\]
Here, $B_1$ stands for the $\mu$-component in $\delta^{18}$O and $B_2$ for the $\mu$-component in $\delta^{13}$C. We have for $s<t$ that
\begin{align*}
& \cov[B_1(t)-B_1(s),B_2(t)-B_2(s)] = \E[(B_1(t)-B_1(s))(B_2(t)-B_2(s))] \\
& \quad = \E[(B_1(t)-B_1(s))(\rho(B_1(t)-B_1(s))+\sqrt{1-\rho^2}(B_3(t)-B_3(s)))]\\
& \quad = \rho \E[(B_1(t)-B_1(s))^2] = \rho(t-s),
\end{align*}
because of the independence of $B_1$ and $B_3$. More generally, for $s<t<u<v$,
\begin{align*}
& \cov[B_1(v)-B_1(t),B_2(u)-B_2(s)] = \rho \E[(B_1(v)-B_1(t))(B_1(u)-B_1(s))] \\
& \quad = \rho \E\left[\left((B_1(v)-B_1(u)) + (B_1(u)-B_1(t))\right)\left((B_1(u)-B_1(t)) + (B_1(t)-B_1(s))\right)\right]\\
& \quad = \rho \E\left[(B_1(u)-B_1(t))^2\right] = \rho (u-t),
\end{align*}
because
\begin{align*}
& \E[(B_1(v)-B_1(u))(B_1(u)-B_1(t))]=\E[(B_1(v)-B_1(u))(B_1(t)-B_1(s))]\\
& \quad =\E[(B_1(u)-B_1(t))(B_1(t)-B_1(s))]=0.
\end{align*}
Thus, the covariance of two increments of $\mu^{\delta^{18}O}$ and $\mu^{\delta^{13}C}$ is modelled as
\[
\E[(\mu^{\delta^{18}O}_{v}-\mu^{\delta^{18}O}_{t})(\mu^{\delta^{13}C}_{u}-\mu^{\delta^{13}C}_{s})] = \rho\; \text{length}((t,v)\cap (s,u)).
\] 
In all cases in the data, either $s=t$ or $u=v$, and therefore $\text{length}((t,v)\cap (s,u))=\min\{v-t,u-s\}$.

A simple bivariate random walk plus noise model for the two series that extends Equation \eqref{E:RWN} consists of the measurement equations
\begin{align*}
y^{\delta^{18}O}_{n} &= \mu^{\delta^{18}O}_{t_n} + \eps^{\delta^{18}O}_{n}, & \eps^{\delta^{18}O}_{n}\stackrel{i.i.d.}{\sim} \mathsf{N}(0,\sigma_{\eps,\delta^{18}O}^2),\\
y^{\delta^{13}C}_{n} &= \mu^{\delta^{13}C}_{t_n} + \eps^{\delta^{13}C}_{n}, & \eps^{\delta^{13}C}_{n}\stackrel{i.i.d.}{\sim} \mathsf{N}(0,\sigma_{\eps,\delta^{13}C}^2),
\end{align*}
where $\eps_t^{\delta^{18}O}$ and $\eps_t^{\delta^{13}C}$ are independent. The interesting interaction of the bivariate specification plays out in the covariance matrix of the transition equation, where we allow for correlation of the transition error from the unobserved component in $\delta^{18}$O with the one in $\delta^{13}$C: 
\begin{align*}
\mu_{t_n+\Delta t_n}^{\delta^{18}O} &= \mu_{t_n}^{\delta^{18}O} + \eta_{t_n}^{\delta^{18}O},\\
\mu_{t_n+\Delta t_n}^{\delta^{13}C} &= \mu_{t_n}^{\delta^{13}C} + \eta_{t_n}^{\delta^{13}C},
\end{align*}
where the state disturbances can be collected in the random vector
$\eta_{t_n} = (\eta_{t_n}^{\delta^{18}O}, \eta_{t_n}^{\delta^{13}C})^T$ that follows a  bivariate normal distribution given by
\begin{equation}
\eta_{t_n} \stackrel{i.i.d.}{\sim} \mathsf{N}(0, Q \Delta {t_n}),\qquad 
Q = \left[\begin{array}{cc}
\sigma^2_{\eta,\delta^{18}O} & \rho \sigma_{\eta,\delta^{18}O}\sigma_{\eta,\delta^{13}C} \\
\rho \sigma_{\eta,\delta^{18}O}\sigma_{\eta,\delta^{13}C} & \sigma^2_{\eta,\delta^{13}C}
\end{array}\right].\label{E:BivCov}
\end{equation}
where $-1\ge \rho \ge 1$ is the correlation coefficient. In this basic specification, we have five
unknown parameters: two measurement error variances, two transition error variances, and the correlation $\rho$.
Given that the Kalman filter and smoothing recursions can be generalized to multivariate model equations \citep[Chapter 4]{Durbin2012}, the methodology discussed in
Section \ref{S:RWN} can still be adopted.

\subsection*{Revisiting the analyses of $\delta^{13}$C and $\delta^{18}$O}

The estimation results of the model are reported in Table \ref{T:BivRWN}. The results are qualitatively very close to Table \ref{T:RWN}. The correlation parameter is estimated at -0.17, and even though the corresponding standard error is much smaller in magnitude, the log-likelihood is close to the sum of the two log-likelihoods from the separate estimation. A plot of the smoothed values and residuals can visually not be distinguished from Figure \ref{F:RWN}, and so it is omitted here.

\begin{table}
\caption{\footnotesize Maximum likelihood parameter estimates for the simplest bivariate random-walk-plus-noise model. Standard errors in parentheses. BIC: Bayes information criterion, LLH: log-likelihood. The lower block reports the independent benchmark in which the two series are modelled separately by the univariate random-walk-plus-noise models of Table~\ref{T:RWN} (i.e.\ $\rho$ fixed to zero): its log-likelihood is the sum of the two univariate values ($7218.57+2992.10$) and its BIC uses four parameters. Allowing $\rho\neq0$ raises the log-likelihood by $19.18$ (a likelihood-ratio statistic of $38.36$ against $\chi^2_1$, $99\%$ critical value $6.63$) and lowers the BIC by $28.4$, so the bivariate coupling is strongly favoured.\label{T:BivRWN}}
\centering
\begin{tabular}{ccc}\hline
 & $\delta^{18}$O & $\delta^{13}$C \\\hline
$\sigma_\eps^2$ & 0.0205 (0.0002) & 0.0341 (0.0004) \\
$\sigma_\eta^2$ & 1.8154 (0.0475) & 1.1991 (0.0396) \\
$\rho$      &   \multicolumn{2}{c}{-0.1700 (0.0209)}  \\\hline
LLH & \multicolumn{2}{c}{10229.85}   \\
BIC & \multicolumn{2}{c}{-20409.34}   \\\hline
\multicolumn{3}{@{}l}{\footnotesize \emph{Independent benchmark} ($\rho\equiv 0$, univariate fits of Table~\ref{T:RWN}):} \\
LLH$_{\rho=0}$ & \multicolumn{2}{c}{10210.67}   \\
BIC$_{\rho=0}$ & \multicolumn{2}{c}{-20380.95}   \\\hline
\end{tabular}
\end{table}

We generalize the bivariate approach to distinguish measurement equation variances according to the ten drill sites introduced in Section~\ref{S:RWN_Sites} and transition equation variances and correlations according to the climate states in Table \ref{T:Climate_States}. That is, there are six different covariance matrices of the type \eqref{E:BivCov} for the different climate states and ten site variances for $\delta^{18}$O and $\delta^{13}$C, respectively, for a total of 38 parameters. The system matrices are spelled out in Appendix \ref{A:Matrices}.

\begin{figure}[h]
\centering
\includegraphics[width=\textwidth]{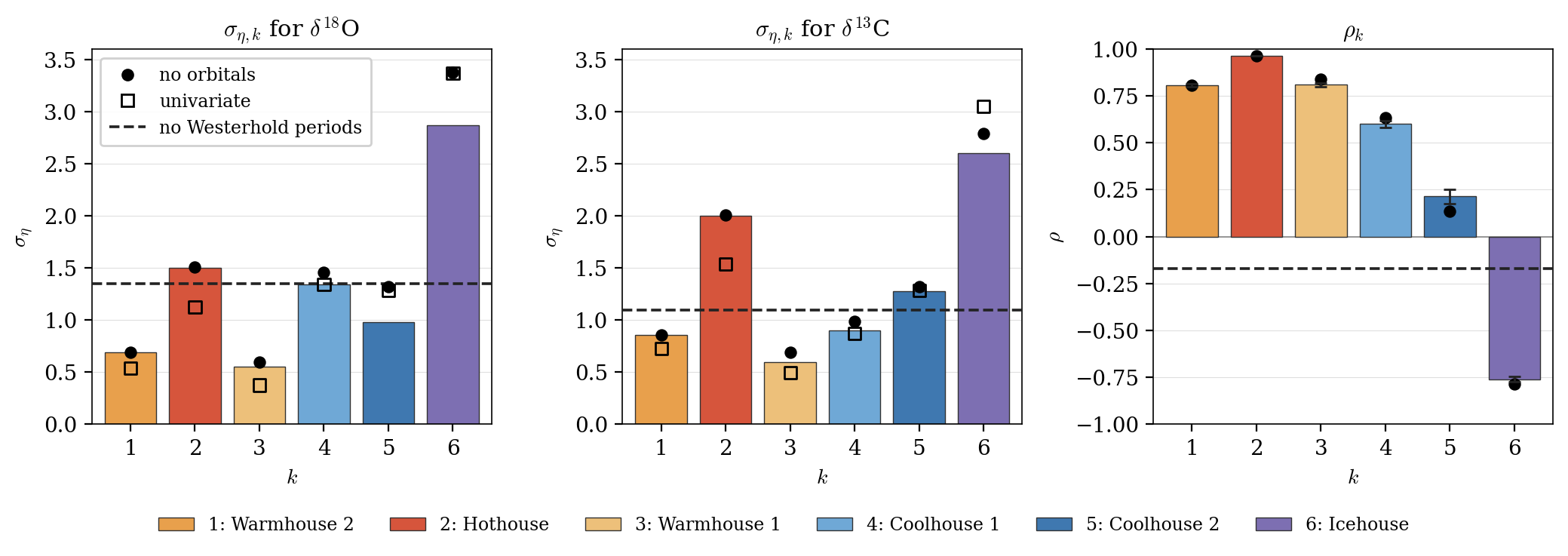}
\caption{\footnotesize Per-climate-state state-equation standard deviations $\sigma_{\eta,k}$ (left and middle panels) and cross-proxy correlations $\rho_k$ (right panel) for three nested bivariate specifications. The $\rho_k$ bars carry $95\%$ confidence whiskers, formed on the transformed scale $2\operatorname{atanh}\rho$ on which the correlation is parameterised and mapped back through $\tanh(\cdot/2)$ so they stay within $[-1,1]$. Coloured bars: the full model with site intercepts, climate-state transition variances and correlations, and period-dependent Milankovitch forcing (Table~\ref{T:Milank_OrbPer}). Black dots (``no orbitals''): the same model without orbital forcing (Table~\ref{T:RWN_SiteBiv}); where a dot lies above its bar, the orbital terms have absorbed part of the stochastic level-innovation variance, an effect concentrated in the cool and icehouse states (4 to 6). Dashed lines (``no Westerhold periods''): the whole-sample constant transition specification without climate-state differentiation (Table~\ref{T:BivRWN}). Open squares in the two $\sigma_{\eta,k}$ panels: the univariate (separately estimated) transition standard deviations from the site+periods random-walk-plus-noise model of Table~\ref{T:RWN_SitePV}; they lie below the bars, showing that the joint model assigns more variation to the signal. The abscissa labels 1 through 6 are the six climate states of Table~\ref{T:Climate_States}. \label{F:BV_corr}}
\end{figure}

The correlation of the state disturbances presents itself at a much higher resolution in Table \ref{T:RWN_SiteBiv} in the appendix compared to Table \ref{T:BivRWN}. The state-dependent variances and correlations from Table~\ref{T:RWN_SiteBiv} are shown as the black dots (``no orbitals'') in Figure~\ref{F:BV_corr}, with the whole-sample constant-transition benchmark of Table~\ref{T:BivRWN} as dashed lines (``no Westerhold periods''); the coloured bars in that figure additionally include the period-dependent Milankovitch forcing introduced in the next subsection (Table~\ref{T:Milank_OrbPer}) and are discussed there. There emerges a pattern that high positive correlation in the early record covering warm- and hothouses weakens in the fourth and fifth climate state covering coolhouses and finally reverses sign in the final, icehouse state. Positive correlation in the innovations of two random walks means that there is a tendency that the series move together in local upward or downward runs, and they move in opposite directions in the presence of negative correlation. This pattern was found earlier in \cite{turner2014pliocene}, for example. Correlation is substantial in all periods except the fifth, and thus joint modeling of $\delta^{18}$O and $\delta^{13}$\text{C} is advantageous in the sense that the estimation of the unobserved states $\mu_{t_n}^{(\delta^{18}\text{O},\delta^{13}\text{C})}$ can benefit from the larger information set. This is reflected in the log-likelihood of the joint model reported in Table~\ref{T:RWN_SiteBiv} (21509.25), which is substantially higher than the sum of the two separate log-likelihood values from the univariate site models in Table~\ref{T:RWN_SitePV} in the appendix (20285.32), at the cost of additional parameters. The variances of the transition equation for the different climate states are higher for the first five periods in the joint model compared to the separate univariate models, meaning that more variation in the data is captured by the unobserved component of interest, the ``signal'', and less variation remains in the ``noise.'' The site intercepts $c_s$ are well identified and substantial: the d18O biases for sites 1218, 1258, and 1263 are at the level of $+0.21$ to $+0.29$\permil\ relative to the reference site 1262; the d13C bias for CearaRiseStack reaches $+0.36$\permil. Joint estimation of biases and dynamics adds 18 parameters to the 38-parameter bivariate sites+periods specification, but the log-likelihood improvement of $594$ points (LR statistic $\approx 1188$ against $\chi^2_{18}$) is far larger than the BIC penalty, so the joint model \eqref{E:BivCov} with biases reported in Table~\ref{T:RWN_SiteBiv} is the preferred bivariate random-walk-plus-noise model.


\subsection*{Milankovitch forcings}

The pacing of Cenozoic climate by variations in Earth's orbit (the eccentricity of the orbit, the obliquity of the rotational axis, and the climatic precession) is among the most firmly established results of paleoclimatology \citep{hays1976,imbrie1984,zachos2001}. The specifications above treat the unobserved $\mu$-components as driven purely by stochastic innovations. We now ask how much of their evolution can be attributed to deterministic orbital forcing, by adding the three orbital variables, taken from the astronomical solution La2004 of \cite{laskar2004}, as exogenous regressors on the increments of the level states $\mu^{\delta^{18}O}$ and $\mu^{\delta^{13}C}$. The orbital variables enter the transition equation through their increments over each inter-observation interval, with six coefficients in all (one per orbital variable and proxy). The level transition becomes
\[
\begin{bmatrix} \mu_{t_n+\Delta t_n}^{\delta^{18}O} \\ \mu_{t_n+\Delta t_n}^{\delta^{13}C} \end{bmatrix}
=
\begin{bmatrix} \mu_{t_n}^{\delta^{18}O} \\ \mu_{t_n}^{\delta^{13}C} \end{bmatrix}
+ B
\begin{bmatrix} \Delta e_n \\ \Delta\varepsilon_n \\ \Delta p_n \end{bmatrix}
+
\begin{bmatrix} \eta_{k,t_n}^{\delta^{18}O} \\ \eta_{k,t_n}^{\delta^{13}C} \end{bmatrix},
\]
where $\Delta e_n$, $\Delta\varepsilon_n$ and $\Delta p_n$ are the increments of eccentricity, obliquity and climatic precession over $[t_n,t_{n+1}]$, $B\in\R^{2\times 3}$ collects the six orbital coefficients, and the identity transition matrix leaves each level a random walk modulated by the deterministic orbital drift; the state-disturbance covariance $Q_{t_n}$ is unchanged (full specification in Appendix~\ref{A:Milank}). All other features of the preferred bivariate model (the site intercepts and measurement variances, and the climate-state transition variances and correlations) are retained.

The three orbital series are taken from the astronomical solution La2004 of \cite{laskar2004}, obtained from the IMCCE / Paris Observatory (file \texttt{INSOLN.LA2004.BTL.100.ASC}).\footnote{\url{https://ssp.imcce.fr/insola/earth/online/earth/La2004/}.} We use eccentricity (dimensionless), obliquity (the tilt of the rotational axis, in radians), and climatic precession, defined as $e\sin\varpi$ with $\varpi$ the longitude of perihelion. Because these quantities are computed by numerical integration of the orbital and rotational dynamics, they are smooth deterministic functions of time, available on a regular $1$-kyr grid from $101$~Ma to the present, which covers the full $67$~Myr span of the benthic record. We interpolate each series with a cubic spline to the calendar age of every observation. The solution is therefore defined at every calendar date, including the ages of missing or sparsely sampled observations; for the regression we use only the increments at the observed ages (Appendix~\ref{A:Milank}).

Table~\ref{T:Milank_BIC} in the appendix reports the model comparison. Adding the six orbital coefficients raises the log-likelihood by $116$ points (likelihood-ratio statistic $231$ against $\chi^2_6$, whose $99\%$ critical value is $16.81$) and lowers the BIC by $171$ points: orbital forcing is a strongly significant feature of the benthic record even after the flexible climate-state dynamics of the preferred model are accounted for. The estimated coefficients in Table~\ref{T:Milank_Orb} attribute the explained variation mainly to eccentricity and obliquity. Both load significantly and negatively on $\delta^{18}$O; because lower $\delta^{18}$O corresponds to warmer, less glaciated conditions, rising eccentricity and obliquity are associated with warming, the classical direction of orbital pacing. Eccentricity also loads strongly on $\delta^{13}$C, linking orbital variation to the carbon cycle \citep{turner2014pliocene}, whereas climatic precession contributes little, consistent with the hemispheric antisymmetry of precessional forcing, which largely cancels in the globally integrated benthic signal.

\begin{table}[htbp]
\caption{\footnotesize Estimated orbital forcing coefficients from the bivariate RWN model with (climate-state-invariant) Milankovitch forcing (62~parameters). Each coefficient multiplies the increment of the orbital variable over an inter-observation interval in the level transition of the respective proxy. Standard errors in parentheses. Significance: ${}^{***}p<0.01$, ${}^{**}p<0.05$.\label{T:Milank_Orb}}
\centering
\begin{tabular}{lrr}\hline
 & $\delta^{18}$O & $\delta^{13}$C \\\hline
Eccentricity       & $-1.754^{***}$ & $-2.225^{***}$ \\
              & $(0.203)$ & $(0.217)$ \\
Obliquity          & $-1.267^{***}$ & $+0.109$ \\
              & $(0.149)$ & $(0.137)$ \\
Clim.\ precession  & $+0.148^{***}$ & $-0.090^{**}$ \\
              & $(0.042)$ & $(0.042)$ \\
\hline
\end{tabular}
\end{table}

A single set of coefficients for the entire 67-Myr record is restrictive, however: the climate response to orbital pacing is widely understood to have strengthened as the Cenozoic cooled into the glaciated late Neogene and Quaternary. We therefore let the orbital coefficients vary across the six climate states, in the same manner as the transition variances and correlations, giving six versions of each coefficient (36 in total). The improvement is large (Table~\ref{T:Milank_BIC} in the appendix, bottom row): the log-likelihood rises by a further $331$ points (LR $662$ against $\chi^2_{30}$, $99\%$ critical value $50.89$) and the BIC falls by another $359$ points.

\begin{figure}[htbp]
\centering
\includegraphics[width=\textwidth]{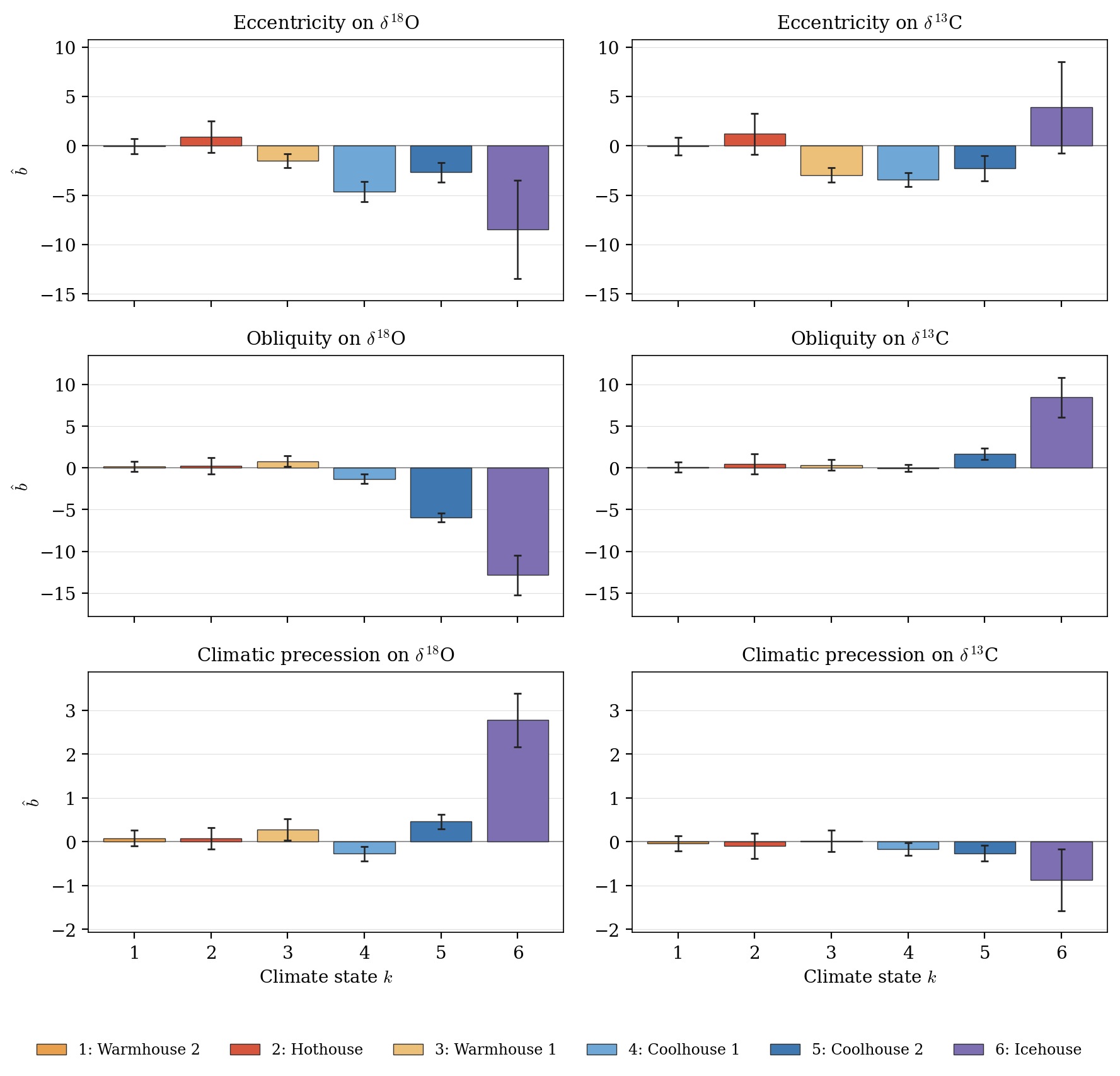}
\caption{\footnotesize Estimated period-dependent Milankovitch coefficients $\hat b$ from the bivariate RWN model (92~parameters), by climate state. Rows are the three orbital variables (eccentricity, obliquity, climatic precession); columns are the two proxies ($\delta^{18}$O, $\delta^{13}$C). Each bar is the coefficient within one climate state, with $\pm 2$ standard-error whiskers; the abscissa labels 1 through 6 are the climate states of Table~\ref{T:Climate_States}. Numerical values are in Table~\ref{T:Milank_OrbPer} in the appendix. \label{F:Milank_periods}}
\end{figure}

Figure~\ref{F:Milank_periods} shows the source of the improvement. In the warm early states (Warmhouse~2 and the Hothouse), the orbital coefficients are small and statistically indistinguishable from zero for both proxies: although the orbital forcing is of course present throughout, there is little detectable orbital response in the benthic record of the early Cenozoic greenhouse. The sparse and unevenly spaced sampling of the early record (Figure~\ref{F:Data}) also limits the orbital frequencies that can be resolved there, so this absence should be read with some caution. The coefficients then grow in magnitude through the Coolhouse states and reach their largest and most significant values in the Icehouse, where obliquity dominates the $\delta^{18}$O response (a coefficient of $-12.9$, $t\approx-11$) and eccentricity and obliquity both load heavily on $\delta^{13}$C. This progression (negligible orbital response in the early greenhouse, strengthening through the cooling Cenozoic, and strongest in the glaciated icehouse) mirrors the established picture of orbital control over late-Cenozoic climate and emerges here directly from the state space estimation. The full set of 36 coefficients with standard errors is reported in Table~\ref{T:Milank_OrbPer} in the appendix.

The Kalman-smoothed level states from the Milankovitch models remain visually close to those of the preferred RWN, because the orbital terms modulate rather than replace the stochastic dynamics. The orbital structure is therefore summarized by the coefficients in Table~\ref{T:Milank_Orb} and Figure~\ref{F:Milank_periods} rather than by the smoothed level itself. Including the orbital forcing does, however, reduce the estimated stochastic transition variances $\sigma^2_{\eta,k}$ in the cool and icehouse states, where the orbital coefficients are largest. This is visible in Figure~\ref{F:BV_corr}: the coloured bars are the present period-dependent Milankovitch model, and the ``no orbitals'' dots, which are the same model without the orbital term, lie above the bars for climate states 4 to 6, indicating that part of the stochastic signal variation in the cooling Cenozoic is absorbed by the deterministic orbital component.

\begin{figure}[htbp]
\centering
\includegraphics[width=\textwidth]{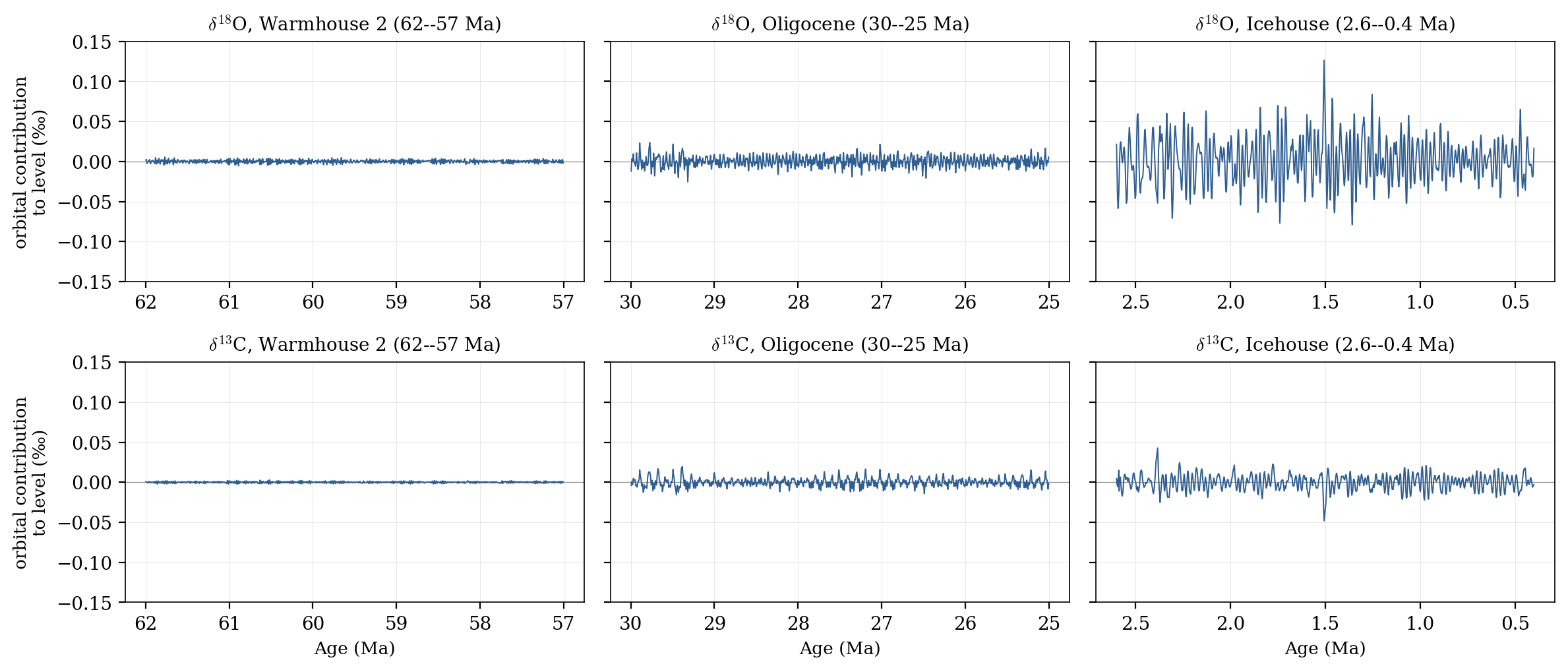}
\caption{\footnotesize Orbital contribution to the smoothed level: the difference between the Kalman-smoothed level of the period-dependent Milankovitch model with and without the orbital term, with the estimated variances held fixed, for $\delta^{18}$O (top row) and $\delta^{13}$C (bottom row) in three windows (columns). All panels share the same vertical scale. The contribution is negligible in the Warmhouse~2 greenhouse, where the orbital coefficients are indistinguishable from zero; it is eccentricity-paced and of order $0.02$\permil\ in the Oligocene; and it is obliquity-paced, reaching $\pm 0.13$\permil\ in $\delta^{18}$O, in the Icehouse. The abscissa is age in millions of years before present.\label{F:OrbContrib}}
\end{figure}

Figure~\ref{F:OrbContrib} makes the orbital imprint on the extracted level explicit, plotting the difference between the smoothed level with and without the orbital term (the estimated variances held fixed) in three windows. The contribution scales with the orbital coefficients, not with the sampling density: it is negligible in the early greenhouse (below $0.01$\permil), even though the inter-observation gaps are largest there; it is eccentricity-paced and of order $0.02$\permil\ in the Oligocene; and it reaches $\pm 0.13$\permil\ in $\delta^{18}$O, with a clear obliquity ($\sim 41$~kyr) pacing, in the Icehouse. Even at its largest it is small against the one-to-three per mil range of the record, so the smoothed level stays visually close to that of the preferred RWN: the orbital signal is expressed mainly in the reduced transition variance rather than in a shift of the level.

A further benefit of the orbital extension is that it sharpens the model's use as an interpolator. The Kalman smoother reconstructs the latent signal $\mu_t$ at any age, including ages at which one or both proxies are unobserved, by treating those points as missing observations. Because the La2004 solution is a deterministic function of time, the orbital increments $\Delta e_n$, $\Delta\varepsilon_n$, $\Delta p_n$ (Appendix~\ref{A:Milank}) are known at every such age as well, so the orbital term continues to drive the level increments where no proxy data are available. The Milankovitch cycles therefore help constrain the reconstruction at unobserved ages, rather than leaving a featureless stochastic interpolation between the data points. Moreover, since the orbital series can be evaluated on any grid, the fitted model can be applied, with its already estimated parameters and without any re-estimation, to a user-specified set of ages: running the filter and smoother on the union of the observed times and, for instance, a regularly spaced grid yields an equidistantly time-stamped reconstruction of $\mu^{\delta^{18}O}_t$ and $\mu^{\delta^{13}C}_t$, with pointwise standard errors, as a side benefit of the analysis.

\subsection*{Bivariate integrated random walk plus noise}

The IWN extension of Section~\ref{S:IWN} carries over to the bivariate framework as a
smoothing illustration. We use the pure $m$-fold form with $m=2$, in which the only state
innovation for each proxy enters at the slope (highest-order) derivative; there is no
separate level innovation. The state vector is $\alpha_t = (\mu_t^{\delta^{18}\text{O}},\,
\beta_t^{\delta^{18}\text{O}},\,\allowbreak \mu_t^{\delta^{13}\text{C}},\, \beta_t^{\delta^{13}\text{C}})^T$,
where $\beta_t$ denotes the slope. Given the slope, the level evolves deterministically over a
step, $\mu_{t_n+\Delta t_n}^{(\cdot)} = \mu_{t_n}^{(\cdot)} + \Delta t_n\,\beta_{t_n}^{(\cdot)}$,
while the innovations enter the slopes, with the cross-proxy dependence acting on the slope
innovations:
\[
\beta_{t_n+\Delta t_n}^{(\cdot)} = \beta_{t_n}^{(\cdot)} + \eta_{t_n}^{\beta,(\cdot)},
\qquad
\begin{pmatrix}\eta_{t_n}^{\beta,\delta^{18}\text{O}}\\[2pt] \eta_{t_n}^{\beta,\delta^{13}\text{C}}\end{pmatrix}
\sim \mathsf{N}\!\left(0,\; \Delta t_n
\begin{pmatrix} \sigma_{\eta,2,k}^{2,\,\delta^{18}\text{O}} & \rho_k\,\sigma_{\eta,2,k}^{\delta^{18}\text{O}}\sigma_{\eta,2,k}^{\delta^{13}\text{C}} \\[2pt]
\cdot & \sigma_{\eta,2,k}^{2,\,\delta^{13}\text{C}} \end{pmatrix}\right),
\]
where $(\cdot)$ stands for $\delta^{18}$O or $\delta^{13}$C and $k$ indexes the climate state.
In the pure $m$-fold IWN the white-noise innovation enters only at the highest ($m$-th)
derivative, so the level itself carries no innovation of its own. The cross-proxy correlation
$\rho_k$ therefore acts on the slope innovations rather than on the level innovations as in the
RWN, and the pure IWN cannot reproduce the level-innovation correlation structure of the RWN.
The RWN therefore remains the preferred specification, and the IWN(2) is included only to
illustrate the additional smoothing afforded by higher-order integration. We consider $m=2$ only.

Table~\ref{T:BivIWN_SitePV} in the appendix reports the per-period slope-innovation variances
$\hat\sigma_{\eta,2,k}^2$, the implied per-step level-innovation standard deviation
$\sqrt{\hat Q_{00}}$, and the slope-innovation correlations $\hat\rho_k$, estimated jointly
with site-specific intercepts and measurement variances (56 parameters). The slope
correlations recover the same climate-state pattern as the RWN level correlations in
Table~\ref{T:RWN_SiteBiv}: strong positive dependence in the warm- and hothouse periods,
weakening through the coolhouses, and reversing to strongly negative in the icehouse.

Figure~\ref{F:BivIWN2_d18O} in the appendix displays the Kalman-smoothed level
with pointwise 95\% confidence bands and the corresponding standardized prediction residuals
for both proxies. The IWN(2) curves follow the data less tightly than the RWN counterpart, as
expected from the higher filter order.


\section{Conclusion}\label{S:Concl}

We specified state space models for the time series of $\delta^{18}$O and $\delta^{13}$C provided in \cite{westerhold2020}. The models addressed the central challenges of the time series: non-stationarity with trends in varying directions, irregular spacing of the time stamps of the observations, multiple observations at certain time stamps, different studies of origin of the data, different climate states covered by the data, and covariation in the two series. The models were specified and estimated by maximum likelihood in the time domain, allowing for a direct interpretation as Butterworth filters in the frequency domain. We showed that accounting for the different studies of origin of the data by way of differentiating measurement equation variances substantially improves the model fit. Another improvement is achieved by differentiating the variances in the state equation by climate states, as they were identified in \cite{westerhold2020}. Bivariate specifications are fruitful in that they use the covariation in the $\delta^{18}$O and $\delta^{13}$C series to inform the estimation of the models. We showed that integration orders of the unobserved component, which at the same time are the orders of the corresponding Butterworth filters, up to six are meaningful for the two time series. They yield visually distinct estimated latent components with different degrees of smoothing. Therefore, the order is a design choice made by the researcher, reflecting their preference for a degree of smoothing. Given this choice, and the data, the maximum likelihood estimation yields optimal estimates of the measurement and state equation variances. These correspond to optimal signal-to-noise ratios (and thus cutoff frequencies) for the corresponding Butterworth filters. In summary, well developed statistical theory, both in the time and in the frequency domains, can be employed to specify time series models for $\delta^{18}$O and $\delta^{13}$C that can resolve many of the rich features of these fascinating data. As an important byproduct, state space models offer a way of dealing with the unequal spacing of paleoclimate observations in time by producing equidistant filtered values. In future research, we will address uncertainty in the time stamps of the observations and relate the $\delta^{18}$O and $\delta^{13}$C measurements to paleo-estimates of atmospheric CO$_2$ concentrations.

\clearpage

\appendix

\section{Tables}\label{A:Tables}

\vspace*{\fill}\vspace*{4cm}
\begin{center}
\raisebox{-0.5\height}{\rotatebox{90}{%
\begin{minipage}{0.72\textheight}
\refstepcounter{table}\label{T:RWN_SitV}%
{\normalfont\fontsize{8bp}{9.5bp}\selectfont\raggedright {\bfseries Table~\thetable}\hskip2mm \footnotesize Maximum likelihood parameter estimates for model \eqref{E:RWN_SitV} with site-specific constant intercepts $c_s$. Standard errors in parentheses. Site 1262 is the reference site ($c_{s_0}=0$). Sites refer to the entry in column ``Record'' of the data file \texttt{aba6853\_tables\_s8\_s34.xlsx} of \cite{westerhold2020}. BIC: Bayes information criterion.\par}
\vskip4pt
\centering
\scriptsize
\begin{tabular}{lrccrcc}\hline
Site & No. obs. & $c_s$ ($\delta^{18}$O) & $\sigma_{\eps,s}^2$ ($\delta^{18}$O) & No. obs. & $c_s$ ($\delta^{13}$C) & $\sigma_{\eps,s}^2$ ($\delta^{13}$C) \\\hline
  Site 1262               &  3905 & 0 (ref) & 0.0081 (0.0003) &  3905 & 0 (ref) & 0.0071 (0.0002) \\
  Site 1263               &  3227 & +0.2089 (0.0061) & 0.0116 (0.0004) &  3227 & +0.0112 (0.0060) & 0.0134 (0.0005) \\
  Site 1258               &   301 & +0.2748 (0.0881) & 0.0122 (0.0016) &   301 & -0.0266 (0.0881) & 0.0164 (0.0019) \\
  Site 1218               &  1836 & +0.2496 (0.0986) & 0.0154 (0.0006) &  1836 & +0.2301 (0.0890) & 0.0118 (0.0005) \\
  Site 1265               &   143 & +0.0984 (0.1584) & 0.0122 (0.0020) &   143 & +0.1215 (0.1397) & 0.0066 (0.0011) \\
  Site 1264               &  3960 & +0.1455 (0.1570) & 0.0115 (0.0003) &  3960 & +0.2115 (0.1386) & 0.0112 (0.0003) \\
  Site U1338              &  1370 & +0.0977 (0.1784) & 0.0143 (0.0007) &  1370 & +0.1140 (0.1599) & 0.0084 (0.0004) \\
  Site 1146               &  1987 & +0.1171 (0.1784) & 0.0064 (0.0003) &  1987 & +0.1770 (0.1601) & 0.0076 (0.0003) \\
  Site U1337              &  2633 & +0.1076 (0.1592) & 0.0127 (0.0005) &  2633 & +0.0785 (0.1415) & 0.0183 (0.0006) \\
  Site CearaRiseStack     &  4897 & +0.1112 (0.1876) & 0.0554 (0.0013) &  4577 & +0.3774 (0.1815) & 0.1291 (0.0029) \\
$\sigma_{\eta}^2$ & & & 1.6618 (0.0485) & & & 1.1609 (0.0348) \\\hline
log-likelihood & 24259 & & 9858.93 & 23939 & & 9111.16 \\
BIC & & & -19516.38 & & & -18020.84 \\\hline
\end{tabular}
\end{minipage}%
}}
\end{center}
\vspace*{\fill}
\clearpage

\begin{sidewaystable}
\caption{\footnotesize Maximum likelihood parameter estimates for the RWN model with site-specific constant intercepts $c_s$, site-specific measurement variances $\sigma_{\eps,s}^2$, and period-specific transition variances $\sigma_{\eta,k}^2$. Standard errors in parentheses. Site 1262 is the reference site ($c_{s_0}=0$). Sites refer to the entry in column ``Record'' in the data file \texttt{aba6853\_tables\_s8\_s34.xlsx} of \cite{westerhold2020}. BIC: Bayes information criterion.\label{T:RWN_SitePV}}
\centering
\scriptsize
\begin{tabular}{lrccrcc}\hline
Site / Period & No. obs. & $c_s$ ($\delta^{18}$O) & $\sigma_{\eps,s}^2$ ($\delta^{18}$O) & No. obs. & $c_s$ ($\delta^{13}$C) & $\sigma_{\eps,s}^2$ ($\delta^{13}$C) \\\hline
  Site 1262               &  3905 & 0 (ref) & 0.0098 (0.0003) &  3905 & 0 (ref) & 0.0072 (0.0002) \\
  Site 1263               &  3227 & +0.2084 (0.0065) & 0.0135 (0.0004) &  3227 & +0.0112 (0.0062) & 0.0135 (0.0005) \\
  Site 1258               &   301 & +0.2758 (0.0868) & 0.0139 (0.0017) &   301 & -0.0112 (0.0944) & 0.0110 (0.0017) \\
  Site 1218               &  1836 & +0.2745 (0.0558) & 0.0155 (0.0007) &  1836 & +0.2546 (0.0620) & 0.0128 (0.0005) \\
  Site 1265               &   143 & +0.1201 (0.1374) & 0.0121 (0.0020) &   143 & +0.1399 (0.1185) & 0.0072 (0.0011) \\
  Site 1264               &  3960 & +0.1672 (0.1358) & 0.0114 (0.0003) &  3960 & +0.2284 (0.1172) & 0.0117 (0.0003) \\
  Site U1338              &  1370 & +0.1098 (0.1609) & 0.0140 (0.0007) &  1370 & +0.1791 (0.1410) & 0.0092 (0.0005) \\
  Site 1146               &  1987 & +0.1296 (0.1610) & 0.0064 (0.0003) &  1987 & +0.2406 (0.1413) & 0.0067 (0.0003) \\
  Site U1337              &  2633 & +0.1286 (0.1384) & 0.0125 (0.0005) &  2633 & +0.0969 (0.1207) & 0.0180 (0.0006) \\
  Site CearaRiseStack     &  4897 & +0.1368 (0.1676) & 0.0435 (0.0010) &  4577 & +0.3929 (0.1695) & 0.1099 (0.0026) \\
$\sigma_{\eta,k}^2$ & & & & & & \\
  67.10113 -- 56 Ma (``Warmhouse 2'') & 2761 & & 0.2863 (0.0291) & 2761 & & 0.5256 (0.0386) \\
  56 -- 47 Ma (``Hothouse'') & 3030 & & 1.2571 (0.0911) & 3030 & & 2.3660 (0.1392) \\
  47 -- 34 Ma (``Warmhouse 1'') & 1786 & & 0.1388 (0.0201) & 1784 & & 0.2405 (0.0291) \\
  34 -- 13.9 Ma (``Coolhouse 1'') & 6669 & & 1.7928 (0.0958) & 6671 & & 0.7528 (0.0502) \\
  13.9 -- 3.3 Ma (``Coolhouse 2'') & 6282 & & 1.6442 (0.0933) & 6139 & & 1.6394 (0.0994) \\
  3.3 -- 0.000564 Ma (``Icehouse'') & 3731 & & 11.3680 (0.7136) & 3554 & & 9.3066 (0.8950) \\\hline
log-likelihood & 24259 & & 10701.27 & 23939 & & 9584.05 \\
BIC & & & -21150.69 & & & -18916.24 \\\hline
\end{tabular}
\end{sidewaystable}

\begin{table}\centering\scriptsize
\caption{\footnotesize Pure $m$-fold IWN measurement-equation estimates for $\delta^{18}$O, with drill site-specific intercepts $c_s$ and variances $\hat\sigma^2_{\eps,s}$, for $m\in\{2,4,6,8\}$. Standard errors in parentheses; the first drill site is the reference ($c=0$). Transition parameters in Table~\ref{T:IWN_Sites_PV}.\label{T:IWN_Sites_d18O}}
\begin{tabular}{lcccc}\hline
Site & $m=2$ & $m=4$ & $m=6$ & $m=8$ \\\hline
$c_s$ & \multicolumn{4}{c}{} \\
  1262 & 0 (ref) & 0 (ref) & 0 (ref) & 0 (ref) \\
  1263 & 0.2111 (0.0069) & 0.2141 (0.0073) & 0.2154 (0.0075) & 0.1993 (0.0082) \\
  1258 & 0.2436 (0.0958) & 0.2125 (0.0930) & 0.2005 (0.0920) & 0.2447 (0.0770) \\
  1218 & 0.2432 (0.0529) & 0.2694 (0.0482) & 0.3388 (0.0462) & 0.3443 (0.0368) \\
  1265 & 0.0820 (0.1353) & 0.2278 (0.0907) & 0.3115 (0.0848) & 0.3160 (0.0784) \\
  1264 & 0.1298 (0.1335) & 0.2749 (0.0882) & 0.3597 (0.0818) & 0.3641 (0.0751) \\
  U1338 & 0.1536 (0.1605) & 0.6134 (0.1044) & 0.6493 (0.0945) & 0.6663 (0.0895) \\
  1146 & 0.1675 (0.1608) & 0.5368 (0.1037) & 0.5388 (0.0924) & 0.5713 (0.0872) \\
  U1337 & 0.1036 (0.1368) & 0.2785 (0.0937) & 0.3523 (0.0859) & 0.3744 (0.0798) \\
  CearaRiseStack & 0.0502 (0.1672) & 0.3494 (0.0987) & 0.4458 (0.0876) & 0.4615 (0.0836) \\
$\hat\sigma^2_{\eps,s}$ & \multicolumn{4}{c}{} \\
  1262 & 0.0118 (0.0003) & 0.0139 (0.0003) & 0.0148 (0.0004) & 0.0173 (0.0004) \\
  1263 & 0.0159 (0.0004) & 0.0175 (0.0005) & 0.0180 (0.0005) & 0.0241 (0.0006) \\
  1258 & 0.0192 (0.0018) & 0.0221 (0.0020) & 0.0228 (0.0020) & 0.0248 (0.0025) \\
  1218 & 0.0199 (0.0007) & 0.0253 (0.0009) & 0.0260 (0.0009) & 0.0268 (0.0009) \\
  1265 & 0.0150 (0.0019) & 0.0196 (0.0024) & 0.0200 (0.0024) & 0.0200 (0.0025) \\
  1264 & 0.0146 (0.0004) & 0.0227 (0.0005) & 0.0234 (0.0005) & 0.0236 (0.0005) \\
  U1338 & 0.0238 (0.0011) & 0.0392 (0.0015) & 0.0439 (0.0017) & 0.0441 (0.0017) \\
  1146 & 0.0093 (0.0004) & 0.0200 (0.0007) & 0.0207 (0.0007) & 0.0206 (0.0007) \\
  U1337 & 0.0179 (0.0006) & 0.0284 (0.0008) & 0.0290 (0.0008) & 0.0292 (0.0008) \\
  CearaRiseStack & 0.0478 (0.0011) & 0.0550 (0.0012) & 0.0583 (0.0012) & 0.0944 (0.0021) \\
\hline
log-likelihood & 9086.51 & 6906.71 & 5650.58 & 4194.15 \\
BIC & -17921.16 & -13561.56 & -11049.31 & -8136.44 \\\hline
\end{tabular}\end{table}

\begin{table}\centering\scriptsize
\caption{\footnotesize Pure $m$-fold IWN measurement-equation estimates for $\delta^{13}$C, with drill site-specific intercepts $c_s$ and variances $\hat\sigma^2_{\eps,s}$, for $m\in\{2,4,6,8\}$. Standard errors in parentheses; the first drill site is the reference ($c=0$). Transition parameters in Table~\ref{T:IWN_Sites_PV}.\label{T:IWN_Sites_d13C}}
\begin{tabular}{lcccc}\hline
Site & $m=2$ & $m=4$ & $m=6$ & $m=8$ \\\hline
$c_s$ & \multicolumn{4}{c}{} \\
  1262 & 0 (ref) & 0 (ref) & 0 (ref) & 0 (ref) \\
  1263 & 0.0160 (0.0066) & 0.0190 (0.0077) & 0.0189 (0.0080) & 0.0192 (0.0081) \\
  1258 & -0.0106 (0.1061) & -0.0386 (0.1054) & -0.0566 (0.1054) & -0.0647 (0.1050) \\
  1218 & 0.2495 (0.0592) & 0.3067 (0.0548) & 0.3615 (0.0416) & 0.4042 (0.0385) \\
  1265 & 0.0940 (0.1135) & 0.2272 (0.0880) & 0.2804 (0.0775) & 0.3205 (0.0739) \\
  1264 & 0.1755 (0.1119) & 0.3037 (0.0861) & 0.3664 (0.0751) & 0.4117 (0.0712) \\
  U1338 & 0.2581 (0.1358) & 0.3848 (0.1041) & 0.4555 (0.0931) & 0.5170 (0.0888) \\
  1146 & 0.3172 (0.1366) & 0.4010 (0.1053) & 0.4538 (0.0936) & 0.5153 (0.0893) \\
  U1337 & 0.0512 (0.1162) & 0.1761 (0.0923) & 0.2506 (0.0817) & 0.2970 (0.0778) \\
  CearaRiseStack & 0.2989 (0.1716) & 0.5539 (0.1185) & 0.6577 (0.1056) & 0.7117 (0.1001) \\
$\hat\sigma^2_{\eps,s}$ & \multicolumn{4}{c}{} \\
  1262 & 0.0095 (0.0003) & 0.0152 (0.0004) & 0.0164 (0.0004) & 0.0172 (0.0004) \\
  1263 & 0.0161 (0.0005) & 0.0185 (0.0005) & 0.0210 (0.0006) & 0.0214 (0.0006) \\
  1258 & 0.0180 (0.0018) & 0.0232 (0.0021) & 0.0251 (0.0023) & 0.0258 (0.0019) \\
  1218 & 0.0157 (0.0006) & 0.0190 (0.0006) & 0.0194 (0.0007) & 0.0198 (0.0007) \\
  1265 & 0.0094 (0.0012) & 0.0130 (0.0016) & 0.0134 (0.0017) & 0.0136 (0.0017) \\
  1264 & 0.0142 (0.0003) & 0.0188 (0.0004) & 0.0193 (0.0004) & 0.0194 (0.0004) \\
  U1338 & 0.0150 (0.0006) & 0.0203 (0.0008) & 0.0211 (0.0008) & 0.0217 (0.0009) \\
  1146 & 0.0099 (0.0004) & 0.0164 (0.0007) & 0.0190 (0.0006) & 0.0196 (0.0006) \\
  U1337 & 0.0217 (0.0007) & 0.0314 (0.0009) & 0.0324 (0.0009) & 0.0328 (0.0009) \\
  CearaRiseStack & 0.1178 (0.0027) & 0.1269 (0.0031) & 0.1700 (0.0036) & 0.1701 (0.0036) \\
\hline
log-likelihood & 8046.04 & 5623.02 & 4404.46 & 3409.00 \\
BIC & -15840.23 & -10994.18 & -8557.06 & -6566.14 \\\hline
\end{tabular}\end{table}

\begin{sidewaystable}
\setlength{\tabcolsep}{4pt}\scriptsize\centering
\caption{\footnotesize Pure $m$-fold IWN transition parameters with measurement variances differentiated by drill sites. For each order $m$ and climate state $k$ we report the continuous-time innovation variance rate $\hat\sigma^2_{\eta,m,k}$, the implied per-step level-innovation standard deviation $\sqrt{\hat Q_{00}}$ (Equation~\eqref{E:Q00}, evaluated at the within-state mean of $\Delta t^{2m-1}$), and the signal-to-noise ratio $\hat q_{m,k}=\hat Q_{00}/\bar\sigma^2_\eps$. The rate $\hat\sigma^2_{\eta}$ grows steeply with $m$ because it carries units $[\,\cdot\,]\,[\text{time}]^{-(2m-1)}$; the dimension-stable $\sqrt{\hat Q_{00}}$ and $\hat q_{m,k}$ decline with $m$, which is the increasing smoothing.\label{T:IWN_Sites_PV}}
\begin{tabular}{llrrrrrrr}\hline
\multicolumn{9}{l}{$\delta^{18}$O} \\
$m$ & quantity & $k=1$ & $k=2$ & $k=3$ & $k=4$ & $k=5$ & $k=6$ & LLH \\\hline
2 & $\hat\sigma^2_{\eta}$ & 200.9089 & 5632.5878 & 12.8510 & 5395.4179 & $2.7\!\times\!10^{4}$ & $5.4\!\times\!10^{5}$ & 9086.5 \\
  & $\sqrt{\hat Q_{00}}$ & 0.0045 & 0.0225 & 0.0032 & 0.0135 & 0.0096 & 0.0297 & \\
  & $\hat q_{m,k}$ & 0.0012 & 0.0299 & $6.2\!\times\!10^{-4}$ & 0.0108 & 0.0054 & 0.0523 & \\[2pt]
4 & $\hat\sigma^2_{\eta}$ & $1.0\!\times\!10^{7}$ & $2.9\!\times\!10^{10}$ & $6.1\!\times\!10^{4}$ & $5.2\!\times\!10^{6}$ & $7.4\!\times\!10^{5}$ & $3.2\!\times\!10^{14}$ & 6906.7 \\
  & $\sqrt{\hat Q_{00}}$ & $1.7\!\times\!10^{-4}$ & 0.0165 & $2.0\!\times\!10^{-4}$ & $5.4\!\times\!10^{-5}$ & $1.5\!\times\!10^{-7}$ & 0.0037 & \\
  & $\hat q_{m,k}$ & $1.4\!\times\!10^{-6}$ & 0.0122 & $1.8\!\times\!10^{-6}$ & $1.3\!\times\!10^{-7}$ & $1.0\!\times\!10^{-12}$ & $6.2\!\times\!10^{-4}$ & \\[2pt]
6 & $\hat\sigma^2_{\eta}$ & $5.3\!\times\!10^{11}$ & $2.0\!\times\!10^{17}$ & $2.5\!\times\!10^{8}$ & $1.6\!\times\!10^{11}$ & $1.3\!\times\!10^{7}$ & $1.2\!\times\!10^{23}$ & 5650.6 \\
  & $\sqrt{\hat Q_{00}}$ & $4.8\!\times\!10^{-6}$ & 0.0074 & $6.5\!\times\!10^{-6}$ & $7.9\!\times\!10^{-7}$ & $1.9\!\times\!10^{-12}$ & $2.4\!\times\!10^{-4}$ & \\
  & $\hat q_{m,k}$ & $9.9\!\times\!10^{-10}$ & 0.0024 & $1.9\!\times\!10^{-9}$ & $2.7\!\times\!10^{-11}$ & $1.6\!\times\!10^{-22}$ & $2.6\!\times\!10^{-6}$ & \\[2pt]
8 & $\hat\sigma^2_{\eta}$ & $3.3\!\times\!10^{16}$ & $1.8\!\times\!10^{20}$ & $5.5\!\times\!10^{8}$ & $5.7\!\times\!10^{15}$ & $4.4\!\times\!10^{11}$ & $1.1\!\times\!10^{26}$ & 4194.1 \\
  & $\sqrt{\hat Q_{00}}$ & $7.7\!\times\!10^{-8}$ & $2.0\!\times\!10^{-5}$ & $2.6\!\times\!10^{-9}$ & $6.8\!\times\!10^{-9}$ & $6.6\!\times\!10^{-16}$ & $1.4\!\times\!10^{-8}$ & \\
  & $\hat q_{m,k}$ & $2.4\!\times\!10^{-13}$ & $1.6\!\times\!10^{-8}$ & $2.9\!\times\!10^{-16}$ & $1.9\!\times\!10^{-15}$ & $1.8\!\times\!10^{-29}$ & $8.2\!\times\!10^{-15}$ & \\[2pt]
\hline
\multicolumn{9}{l}{$\delta^{13}$C} \\
$m$ & quantity & $k=1$ & $k=2$ & $k=3$ & $k=4$ & $k=5$ & $k=6$ & LLH \\\hline
2 & $\hat\sigma^2_{\eta}$ & 1451.1023 & $2.1\!\times\!10^{4}$ & 38.5933 & 950.2696 & $2.0\!\times\!10^{4}$ & $2.0\!\times\!10^{5}$ & 8046.0 \\
  & $\sqrt{\hat Q_{00}}$ & 0.0122 & 0.0438 & 0.0056 & 0.0057 & 0.0085 & 0.0187 & \\
  & $\hat q_{m,k}$ & 0.0097 & 0.1252 & 0.0021 & 0.0021 & 0.0047 & 0.0228 & \\[2pt]
4 & $\hat\sigma^2_{\eta}$ & $1.1\!\times\!10^{7}$ & $2.4\!\times\!10^{11}$ & $5.7\!\times\!10^{5}$ & $5.4\!\times\!10^{6}$ & $2.5\!\times\!10^{8}$ & $2.5\!\times\!10^{13}$ & 5623.0 \\
  & $\sqrt{\hat Q_{00}}$ & $1.9\!\times\!10^{-4}$ & 0.0476 & $6.2\!\times\!10^{-4}$ & $5.4\!\times\!10^{-5}$ & $2.8\!\times\!10^{-6}$ & 0.0011 & \\
  & $\hat q_{m,k}$ & $1.8\!\times\!10^{-6}$ & 0.1200 & $2.0\!\times\!10^{-5}$ & $1.6\!\times\!10^{-7}$ & $4.1\!\times\!10^{-10}$ & $6.2\!\times\!10^{-5}$ & \\[2pt]
6 & $\hat\sigma^2_{\eta}$ & $5.2\!\times\!10^{11}$ & $2.1\!\times\!10^{18}$ & $4.6\!\times\!10^{7}$ & $3.3\!\times\!10^{11}$ & $2.7\!\times\!10^{12}$ & $2.1\!\times\!10^{10}$ & 4404.5 \\
  & $\sqrt{\hat Q_{00}}$ & $4.8\!\times\!10^{-6}$ & 0.0238 & $2.8\!\times\!10^{-6}$ & $1.1\!\times\!10^{-6}$ & $8.7\!\times\!10^{-10}$ & $1.1\!\times\!10^{-10}$ & \\
  & $\hat q_{m,k}$ & $1.1\!\times\!10^{-9}$ & 0.0279 & $4.0\!\times\!10^{-10}$ & $6.2\!\times\!10^{-11}$ & $3.8\!\times\!10^{-17}$ & $5.5\!\times\!10^{-19}$ & \\[2pt]
8 & $\hat\sigma^2_{\eta}$ & $1.8\!\times\!10^{16}$ & $2.9\!\times\!10^{25}$ & $9.9\!\times\!10^{10}$ & $1.9\!\times\!10^{16}$ & $1.7\!\times\!10^{17}$ & $5.0\!\times\!10^{14}$ & 3409.0 \\
  & $\sqrt{\hat Q_{00}}$ & $5.6\!\times\!10^{-8}$ & 0.0079 & $3.5\!\times\!10^{-8}$ & $1.2\!\times\!10^{-8}$ & $4.1\!\times\!10^{-13}$ & $3.0\!\times\!10^{-14}$ & \\
  & $\hat q_{m,k}$ & $1.5\!\times\!10^{-13}$ & 0.0031 & $6.1\!\times\!10^{-14}$ & $7.3\!\times\!10^{-15}$ & $8.2\!\times\!10^{-24}$ & $4.5\!\times\!10^{-26}$ & \\[2pt]
\hline
\end{tabular}\end{sidewaystable}

\begin{sidewaystable}
\caption{\footnotesize Maximum likelihood parameter estimates for the bivariate random-walk-plus-noise model with site-specific constant intercepts $c_s$, site-specific measurement variances $\sigma_{\eps,s}^2$, and period-specific transition variances and correlations. Standard errors in parentheses. Site 1262 is the reference site ($c_{s_0}=0$). Sites refer to the entry in column ``Record'' in the data file \texttt{aba6853\_tables\_s8\_s34.xlsx} of \cite{westerhold2020}. BIC: Bayes information criterion.\label{T:RWN_SiteBiv}}
\centering
\scriptsize
\begin{tabular}{lrccrcc}\hline
Site / Period & No. obs. & $c_s$ ($\delta^{18}$O) & $\sigma_{\eps,s}^2$ ($\delta^{18}$O) & No. obs. & $c_s$ ($\delta^{13}$C) & $\sigma_{\eps,s}^2$ ($\delta^{13}$C) \\\hline
  Site 1262               &  3905 & 0 (ref) & 0.0082 (0.0002) &  3905 & 0 (ref) & 0.0061 (0.0002) \\
  Site 1263               &  3227 & +0.2073 (0.0059) & 0.0108 (0.0004) &  3227 & +0.0129 (0.0057) & 0.0107 (0.0004) \\
  Site 1258               &   301 & +0.2901 (0.0705) & 0.0073 (0.0008) &   301 & -0.0223 (0.0821) & 0.0037 (0.0009) \\
  Site 1218               &  1836 & +0.2449 (0.0622) & 0.0149 (0.0006) &  1836 & +0.2439 (0.0669) & 0.0124 (0.0005) \\
  Site 1265               &   143 & +0.1009 (0.1394) & 0.0116 (0.0019) &   143 & +0.1052 (0.1219) & 0.0070 (0.0011) \\
  Site 1264               &  3960 & +0.1502 (0.1378) & 0.0109 (0.0003) &  3960 & +0.1939 (0.1206) & 0.0113 (0.0003) \\
  Site U1338              &  1370 & +0.0482 (0.1614) & 0.0130 (0.0007) &  1370 & +0.0528 (0.1431) & 0.0084 (0.0005) \\
  Site 1146               &  1987 & +0.0699 (0.1615) & 0.0062 (0.0003) &  1987 & +0.1166 (0.1435) & 0.0064 (0.0003) \\
  Site U1337              &  2633 & +0.1090 (0.1401) & 0.0118 (0.0005) &  2633 & +0.0561 (0.1238) & 0.0172 (0.0006) \\
  Site CearaRiseStack     &  4897 & +0.1185 (0.1701) & 0.0435 (0.0010) &  4577 & +0.3590 (0.1748) & 0.1117 (0.0026) \\
$\sigma_{\eta,k}^2$ & & & & & & \\
  67.10113 -- 56 Ma (``Warmhouse 2'') & 2761 & & 0.4737 (0.0439) & 2761 & & 0.7311 (0.0519) \\
  56 -- 47 Ma (``Hothouse'') & 3030 & & 2.2811 (0.1380) & 3030 & & 4.0256 (0.2164) \\
  47 -- 34 Ma (``Warmhouse 1'') & 1786 & & 0.3487 (0.0426) & 1784 & & 0.4686 (0.0490) \\
  34 -- 13.9 Ma (``Coolhouse 1'') & 6669 & & 2.1111 (0.1106) & 6671 & & 0.9709 (0.0630) \\
  13.9 -- 3.3 Ma (``Coolhouse 2'') & 6282 & & 1.7370 (0.0980) & 6139 & & 1.7340 (0.1043) \\
  3.3 -- 0.000564 Ma (``Icehouse'') & 3731 & & 11.4145 (0.7128) & 3554 & & 7.7973 (0.7007) \\
$\rho_k$ & & & & & & \\
  67.10113 -- 56 Ma (``Warmhouse 2'') & \multicolumn{6}{c}{+0.8071 (0.0211)} \\
  56 -- 47 Ma (``Hothouse'') & \multicolumn{6}{c}{+0.9665 (0.0040)} \\
  47 -- 34 Ma (``Warmhouse 1'') & \multicolumn{6}{c}{+0.8387 (0.0236)} \\
  34 -- 13.9 Ma (``Coolhouse 1'') & \multicolumn{6}{c}{+0.6342 (0.0250)} \\
  13.9 -- 3.3 Ma (``Coolhouse 2'') & \multicolumn{6}{c}{+0.1353 (0.0363)} \\
  3.3 -- 0.000564 Ma (``Icehouse'') & \multicolumn{6}{c}{-0.7869 (0.0306)} \\\hline
log-likelihood & \multicolumn{6}{c}{21509.25} \\
BIC & \multicolumn{6}{c}{-42454.36} \\\hline
\end{tabular}
\end{sidewaystable}

\begin{table}\centering\footnotesize
\caption{\footnotesize Bivariate pure $m$-fold IWN($m=2$): per-period slope-innovation variances $\hat\sigma^2_{\eta,2,k}$ (the sole state driver for each proxy), the implied per-step level-innovation standard deviation $\sqrt{\hat Q_{00}}$ (Equation~\eqref{E:Q00}), and the slope-innovation correlations $\hat\rho_k$ across the two proxies, by climate state. Standard errors in parentheses. Site intercepts $c_s$ and measurement variances $\sigma^2_{\eps,s}$ are estimated jointly (warm-started from the univariate bias-corrected $m=2$ fits) and are omitted. Unlike the RWN, the pure IWN(2) carries no level-innovation correlation; the cross-proxy dependence acts on the slope innovations. $\log$-likelihood $=18194.95$, BIC $=-35825.74$.\label{T:BivIWN_SitePV}}
\begin{tabular}{lrrrrc}\hline
Climate state & $\hat\sigma^2_{\eta,2}$ ($\delta^{18}$O) & $\sqrt{\hat Q_{00}}$ & $\hat\sigma^2_{\eta,2}$ ($\delta^{13}$C) & $\sqrt{\hat Q_{00}}$ & $\hat\rho_k$ \\\hline
Warmhouse 2 & 680.9636 & 0.0084 & 1849.3577 & 0.0138 & 0.9490 (0.0117) \\
Hothouse & $2.1\!\times\!10^{4}$ & 0.0436 & $4.5\!\times\!10^{4}$ & 0.0636 & 0.9979 (0.0005) \\
Warmhouse 1 & 139.3586 & 0.0107 & 268.7225 & 0.0149 & 0.9688 (0.0114) \\
Coolhouse 1 & 4238.0224 & 0.0120 & 1828.4882 & 0.0079 & 0.9748 (0.0052) \\
Coolhouse 2 & $2.3\!\times\!10^{4}$ & 0.0088 & $1.9\!\times\!10^{4}$ & 0.0083 & -0.4521 (0.0682) \\
Icehouse & $5.9\!\times\!10^{5}$ & 0.0312 & $2.9\!\times\!10^{5}$ & 0.0227 & -0.9900 (0.0042) \\
\hline\end{tabular}\end{table}


\begin{table}[htbp]
\caption{\footnotesize Model comparison for the bivariate $(\delta^{18}\text{O},\,\delta^{13}\text{C})$ random-walk-plus-noise model with Milankovitch orbital forcing of the level increments. The preferred bivariate RWN (site intercepts and measurement variances, climate-state transition variances and correlations; 56~parameters) is extended first with six orbital coefficients (eccentricity, obliquity, climatic precession, for each proxy) and then by letting those coefficients vary across the six climate states. $P$: number of parameters. $-\ell$: negative log-likelihood. BIC: Bayes information criterion (smaller preferred). $\Delta$BIC and LR: change relative to the model in the row above; df: added parameters.\label{T:Milank_BIC}}
\centering\footnotesize
\begin{tabular}{lrrrrrr}\hline
Model & $P$ & $-\ell$ & BIC & $\Delta$BIC & LR & df \\\hline
Preferred bivariate RWN (no forcing) & 56 & $-21{,}509.25$ & $-42{,}454.36$ & & & \\
\quad + Milankovitch forcing & 62 & $-21{,}624.94$ & $-42{,}625.29$ & $-170.93$ & $231.38$ & 6 \\
\quad + period-dependent Milank.\ forcing & 92 & $-21{,}955.70$ & $-42{,}984.57$ & $-359.28$ & $661.51$ & 30 \\
\hline
\end{tabular}
\end{table}

\begin{table}[htbp]
\caption{\footnotesize Estimated period-dependent orbital forcing coefficients from the bivariate RWN model with climate-state-specific Milankovitch forcing (92~parameters). Each coefficient multiplies the increment of the orbital variable over an inter-observation interval in the level transition of the respective proxy, separately within each of the six climate states of Table~\ref{T:Climate_States}. Standard errors in parentheses. Significance: ${}^{***}p<0.01$, ${}^{**}p<0.05$.\label{T:Milank_OrbPer}}
\centering\footnotesize
\begin{tabular}{llrr}\hline
Climate state & Orbital variable & $\delta^{18}$O & $\delta^{13}$C \\\hline
Warmhouse 2 (67.1--56 Ma)    & Eccentricity       & $-0.050$ & $-0.033$ \\
                             &                    & $(0.390)$ & $(0.454)$ \\
                             & Obliquity          & $+0.199$ & $+0.119$ \\
                             &                    & $(0.301)$ & $(0.308)$ \\
                             & Clim.\ precession  & $+0.080$ & $-0.042$ \\
                             &                    & $(0.091)$ & $(0.086)$ \\[3pt]
Hothouse (56--47 Ma)         & Eccentricity       & $+0.925$ & $+1.216$ \\
                             &                    & $(0.791)$ & $(1.034)$ \\
                             & Obliquity          & $+0.265$ & $+0.493$ \\
                             &                    & $(0.479)$ & $(0.594)$ \\
                             & Clim.\ precession  & $+0.079$ & $-0.098$ \\
                             &                    & $(0.122)$ & $(0.142)$ \\[3pt]
Warmhouse 1 (47--34 Ma)      & Eccentricity       & $-1.511^{***}$ & $-2.953^{***}$ \\
                             &                    & $(0.342)$ & $(0.357)$ \\
                             & Obliquity          & $+0.813^{**}$ & $+0.360$ \\
                             &                    & $(0.324)$ & $(0.327)$ \\
                             & Clim.\ precession  & $+0.279^{**}$ & $+0.021$ \\
                             &                    & $(0.121)$ & $(0.121)$ \\[3pt]
Coolhouse 1 (34--13.9 Ma)    & Eccentricity       & $-4.624^{***}$ & $-3.414^{***}$ \\
                             &                    & $(0.503)$ & $(0.357)$ \\
                             & Obliquity          & $-1.315^{***}$ & $-0.022$ \\
                             &                    & $(0.277)$ & $(0.219)$ \\
                             & Clim.\ precession  & $-0.273^{***}$ & $-0.164^{**}$ \\
                             &                    & $(0.084)$ & $(0.073)$ \\[3pt]
Coolhouse 2 (13.9--3.3 Ma)   & Eccentricity       & $-2.681^{***}$ & $-2.293^{***}$ \\
                             &                    & $(0.496)$ & $(0.640)$ \\
                             & Obliquity          & $-5.945^{***}$ & $+1.711^{***}$ \\
                             &                    & $(0.281)$ & $(0.349)$ \\
                             & Clim.\ precession  & $+0.463^{***}$ & $-0.263^{***}$ \\
                             &                    & $(0.081)$ & $(0.092)$ \\[3pt]
Icehouse (3.3--0 Ma)         & Eccentricity       & $-8.475^{***}$ & $+3.899$ \\
                             &                    & $(2.499)$ & $(2.324)$ \\
                             & Obliquity          & $-12.866^{***}$ & $+8.470^{***}$ \\
                             &                    & $(1.181)$ & $(1.194)$ \\
                             & Clim.\ precession  & $+2.776^{***}$ & $-0.874^{**}$ \\
                             &                    & $(0.304)$ & $(0.352)$ \\
\hline
\end{tabular}
\end{table}

\clearpage
\section{Figures}\label{A:Figures}

This appendix collects the integrated random-walk-plus-noise (IWN) smoothed-state and residual plots for the $\delta^{18}$O and $\delta^{13}$C series, under the site- and study-based specifications, for integration orders $m=2,4,6,8$.

\begin{figure}[htbp]
\centering
\includegraphics[width=.45\textwidth]{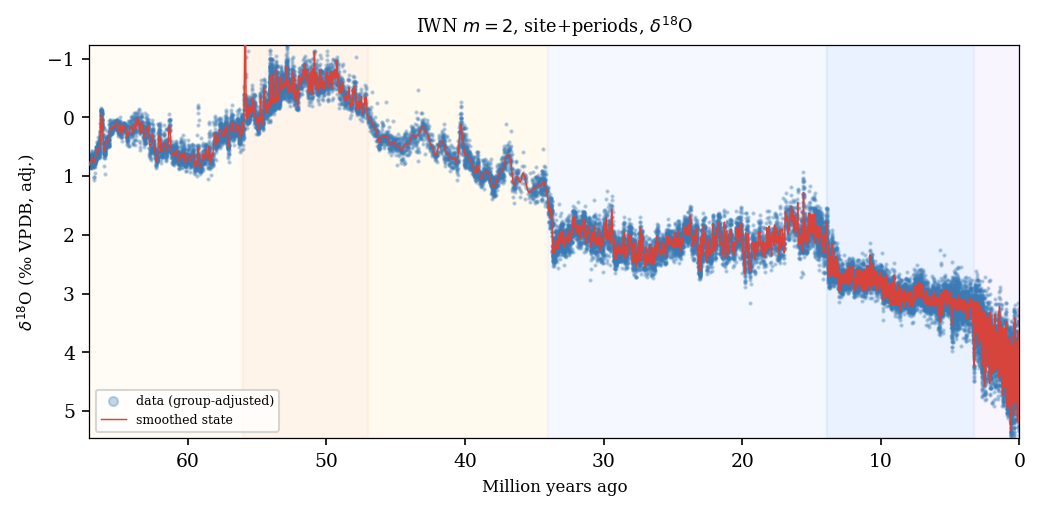}
\includegraphics[width=.45\textwidth]{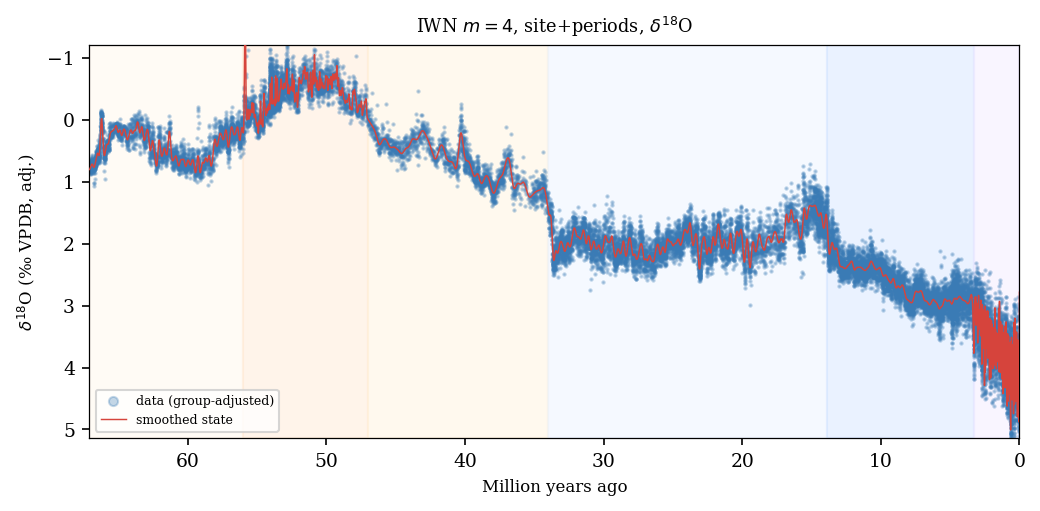}\\
\includegraphics[width=.45\textwidth]{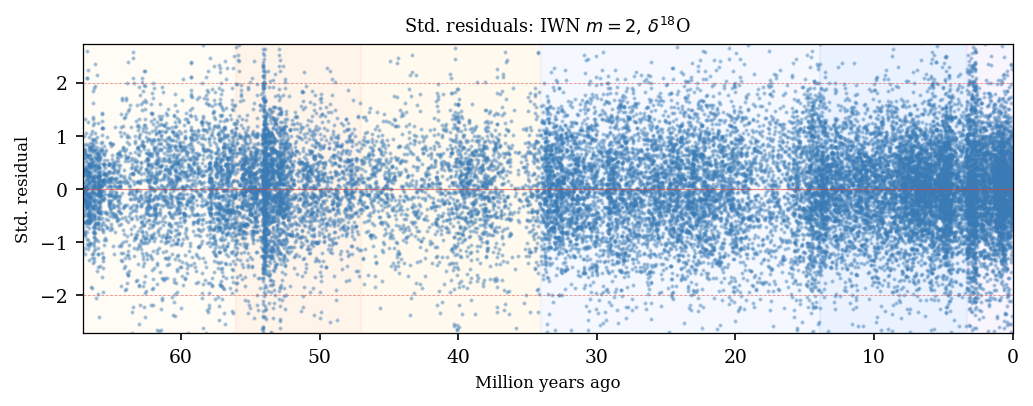}
\includegraphics[width=.45\textwidth]{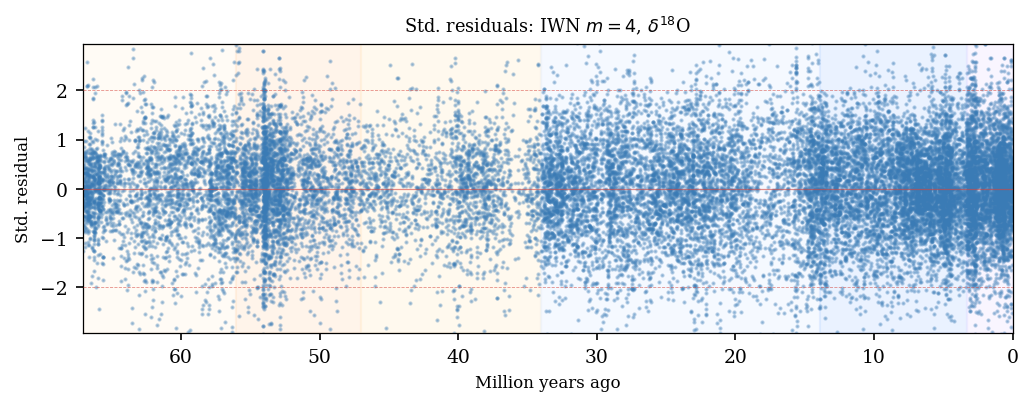}\\[2mm]
\includegraphics[width=.45\textwidth]{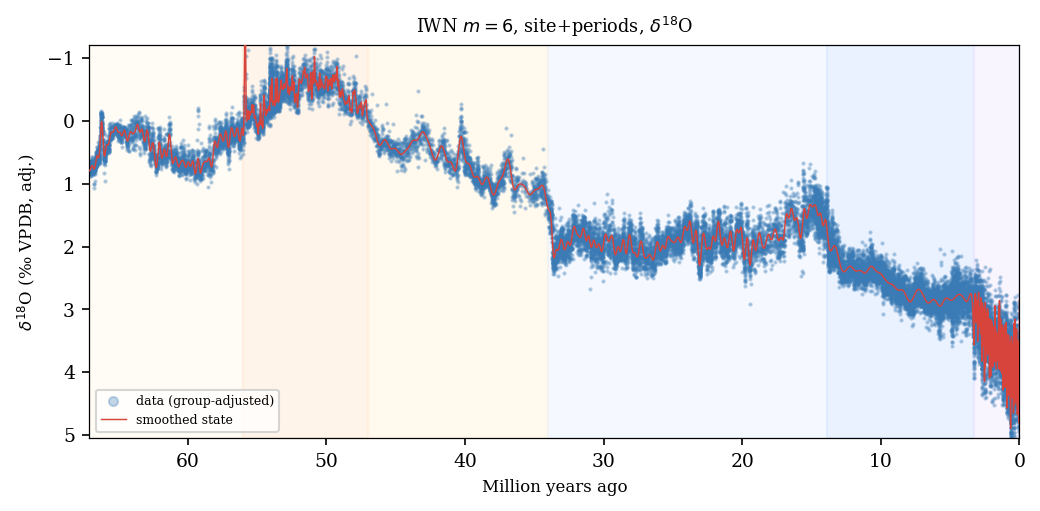}
\includegraphics[width=.45\textwidth]{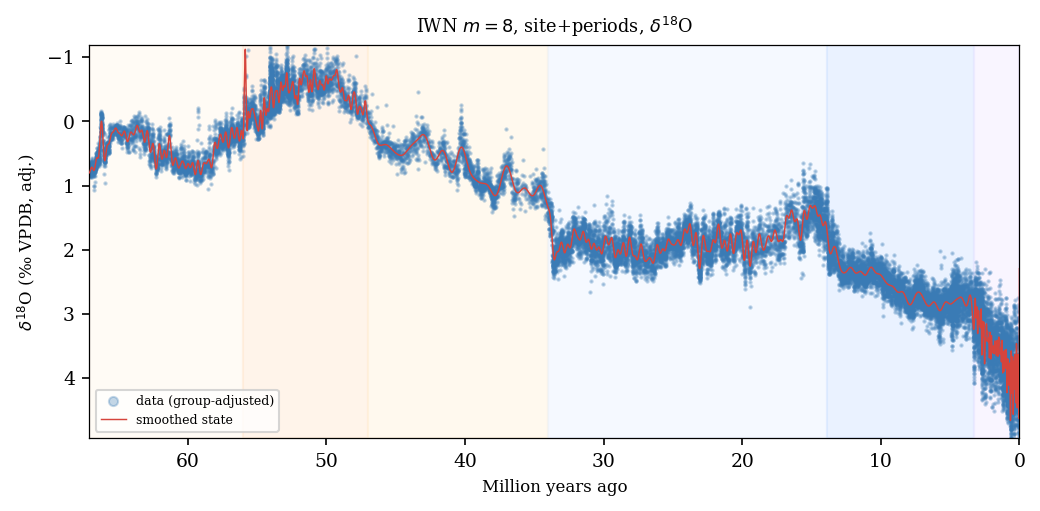}\\
\includegraphics[width=.45\textwidth]{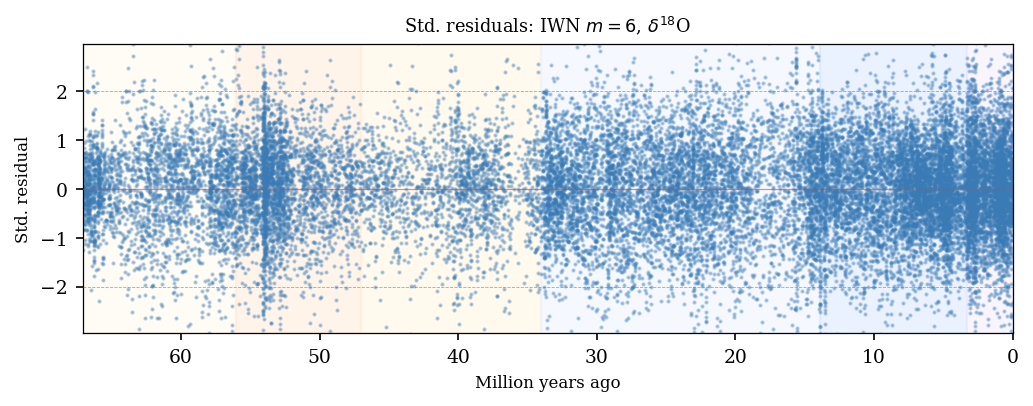}
\includegraphics[width=.45\textwidth]{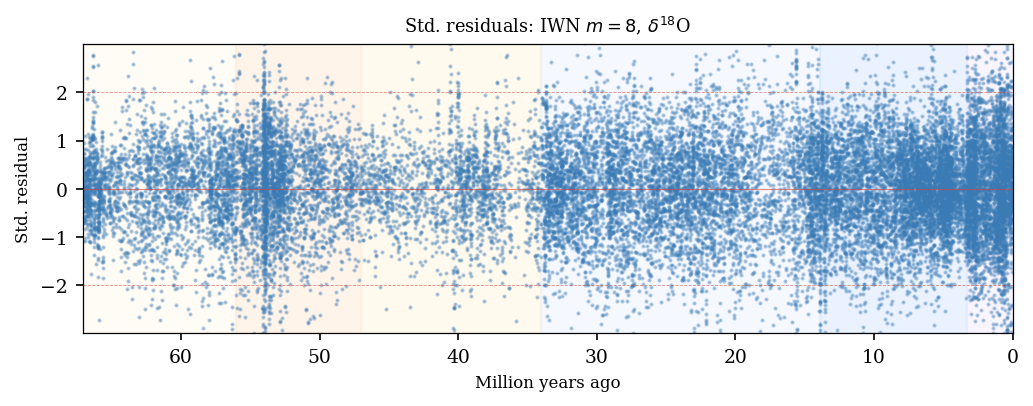}\\
\caption{\footnotesize Site-based specification, $\delta^{18}$O. For each integration order $m$ of model~\eqref{E:IWN_m}, the smoothed state $\E[\mu_{t+\Delta t} | \{y_s,\; s\le t_N\}]$ (red) with the data (blue) is shown on top and the standardized prediction residuals beneath. Upper block: $m=2$ (left) and $m=4$ (right); lower block: $m=6$ (left) and $m=8$ (right). Measurement variances are differentiated by drill site and transition variances by climate state. Note that the $y$-axis for $\delta^{18}$O is reversed, following common practice.\label{F:IRW_Sites_d18O}}
\end{figure}


\begin{figure}[htbp]
\centering
\includegraphics[width=.45\textwidth]{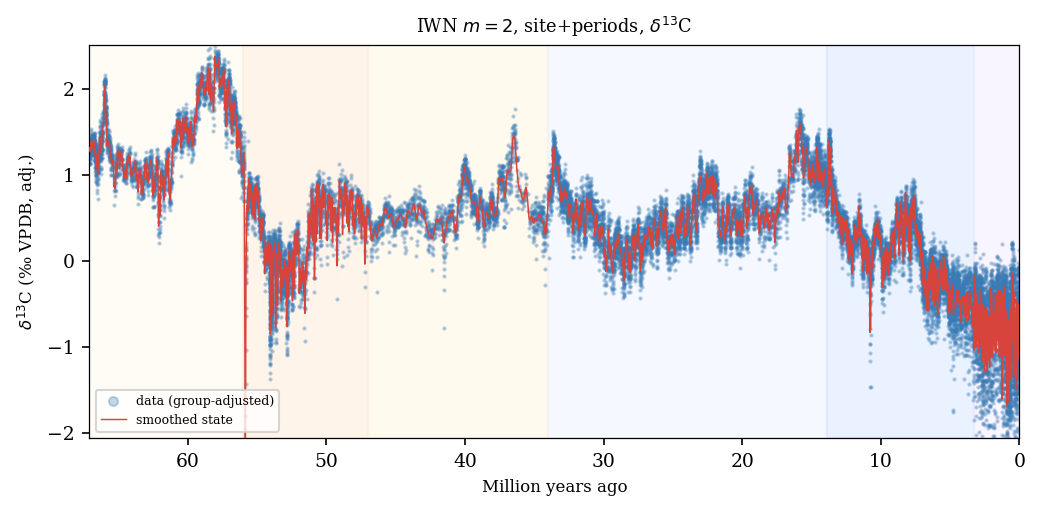}
\includegraphics[width=.45\textwidth]{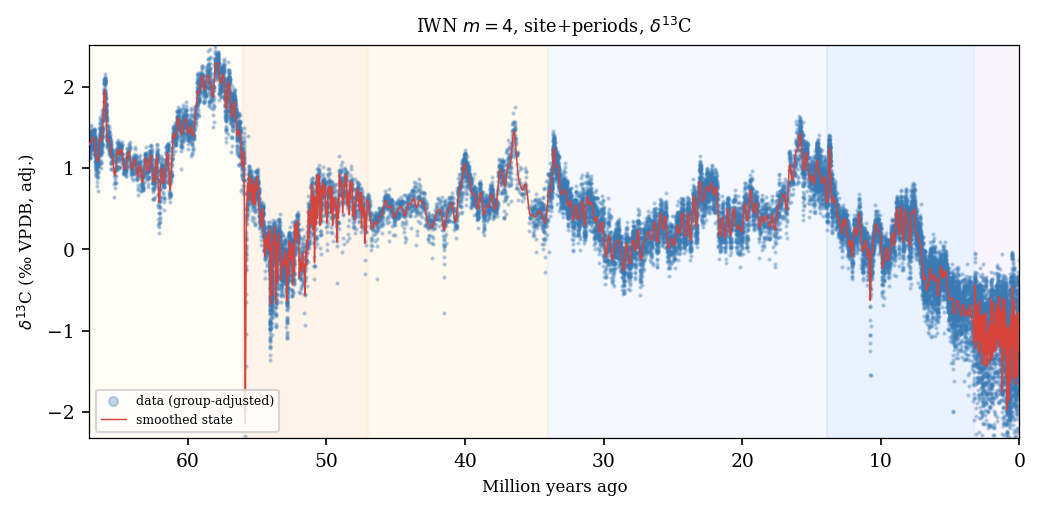}\\
\includegraphics[width=.45\textwidth]{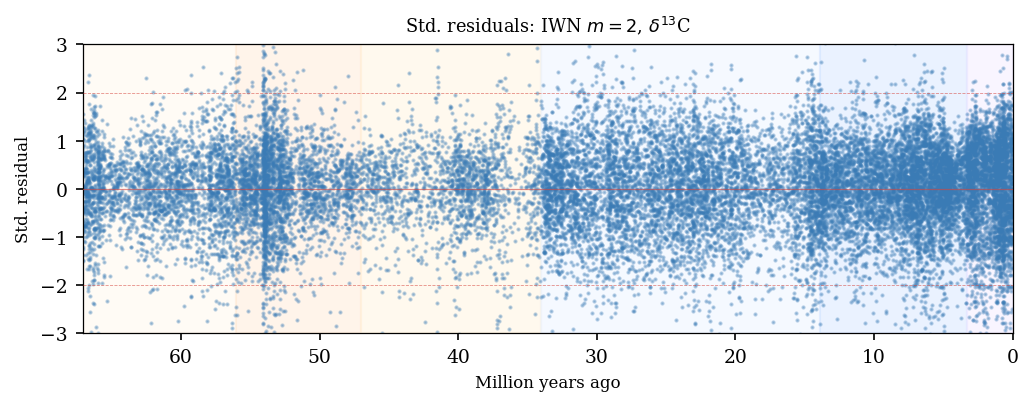}
\includegraphics[width=.45\textwidth]{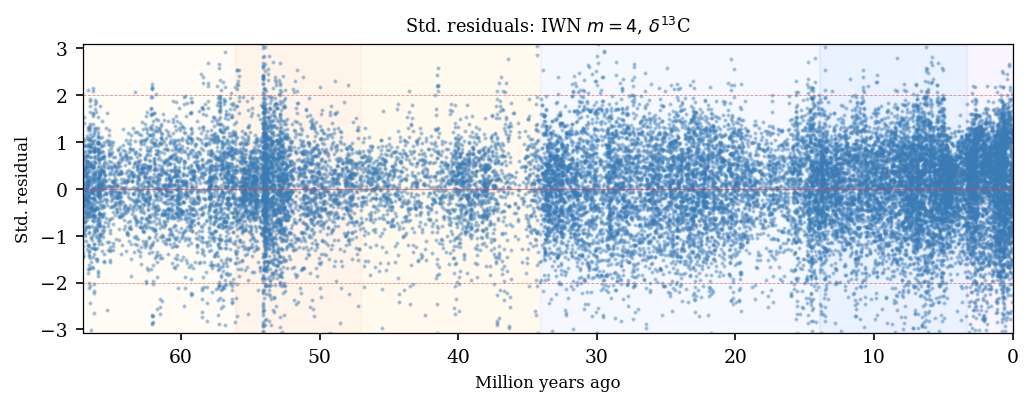}\\[2mm]
\includegraphics[width=.45\textwidth]{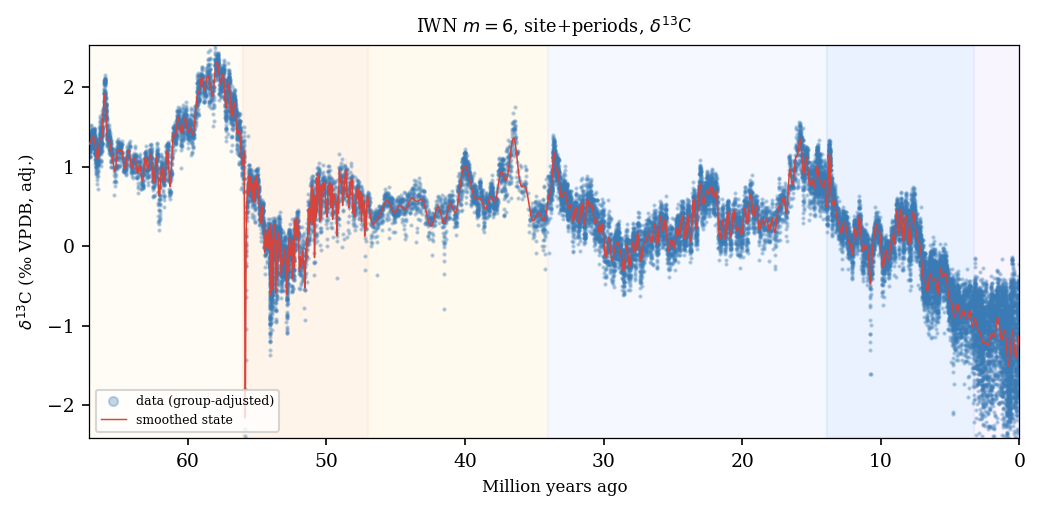}
\includegraphics[width=.45\textwidth]{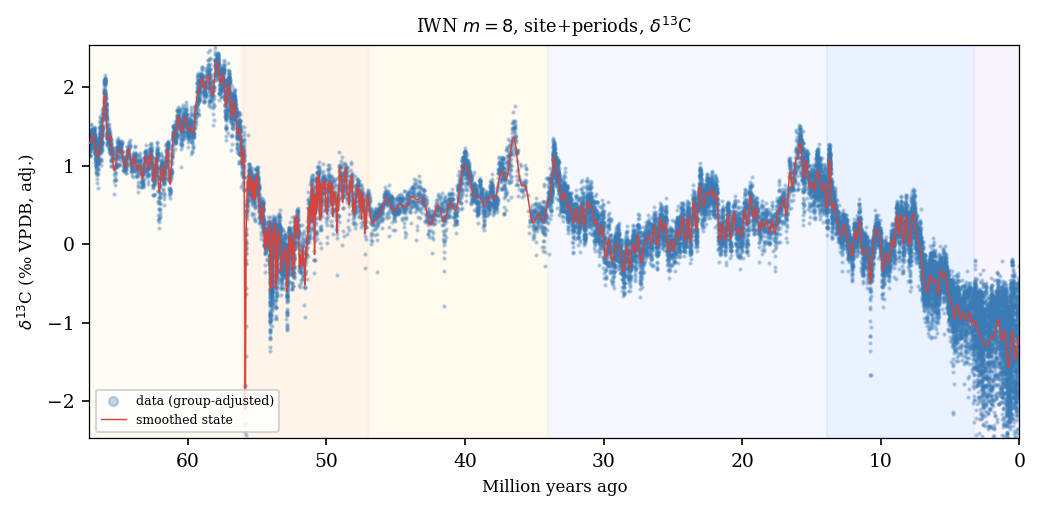}\\
\includegraphics[width=.45\textwidth]{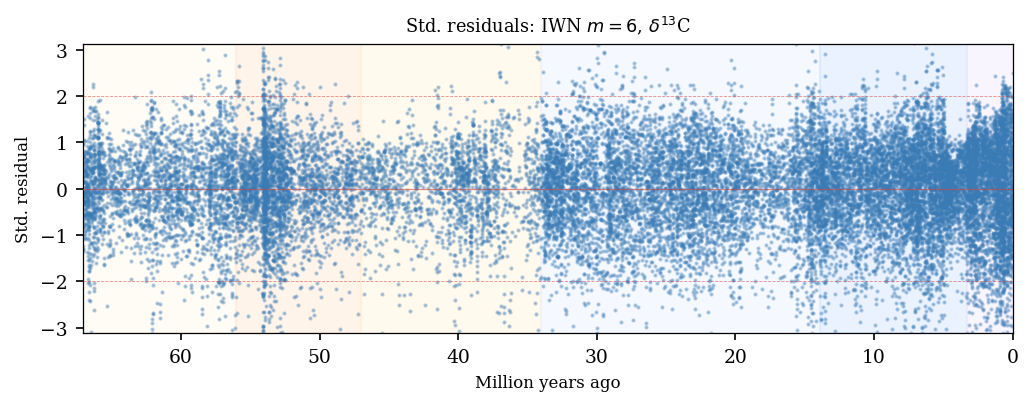}
\includegraphics[width=.45\textwidth]{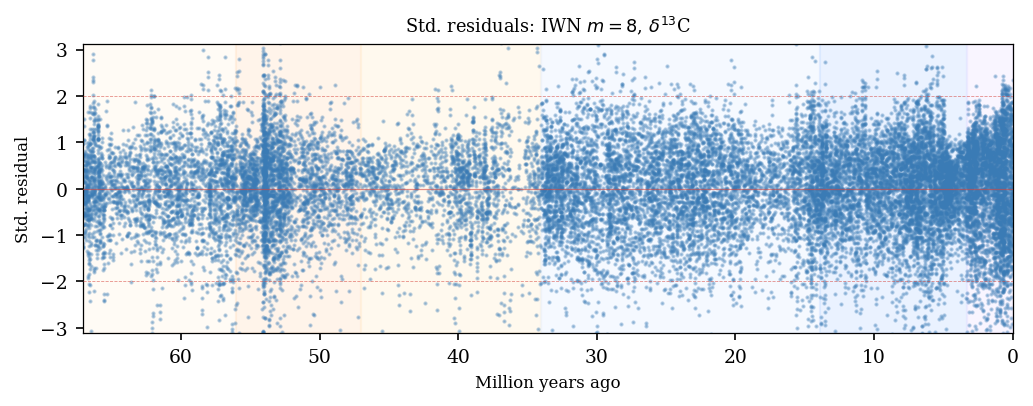}\\
\caption{\footnotesize Site-based specification, $\delta^{13}$C. For each integration order $m$ of model~\eqref{E:IWN_m}, the smoothed state $\E[\mu_{t+\Delta t} | \{y_s,\; s\le t_N\}]$ (red) with the data (blue) is shown on top and the standardized prediction residuals beneath. Upper block: $m=2$ (left) and $m=4$ (right); lower block: $m=6$ (left) and $m=8$ (right). Measurement variances are differentiated by drill site and transition variances by climate state.\label{F:IRW_Sites_d13C}}
\end{figure}


\begin{figure}[htbp]
\centering
\includegraphics[width=.45\textwidth]{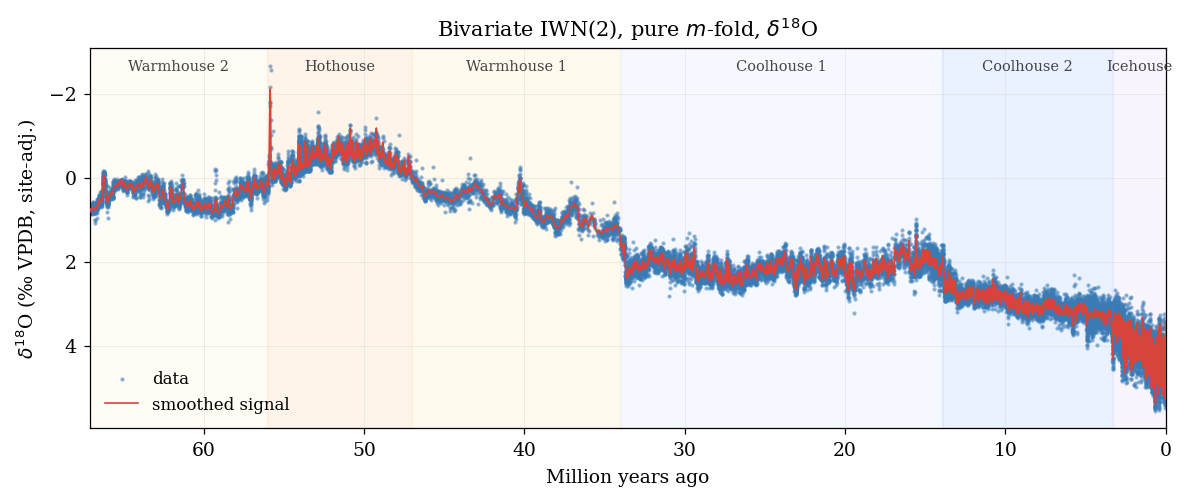}
\includegraphics[width=.45\textwidth]{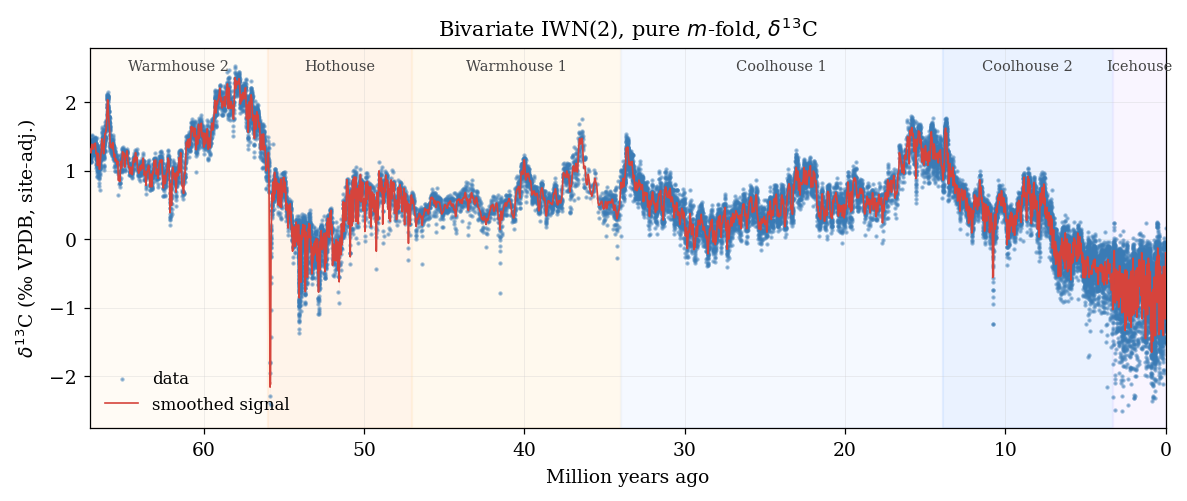}\\
\includegraphics[width=.45\textwidth]{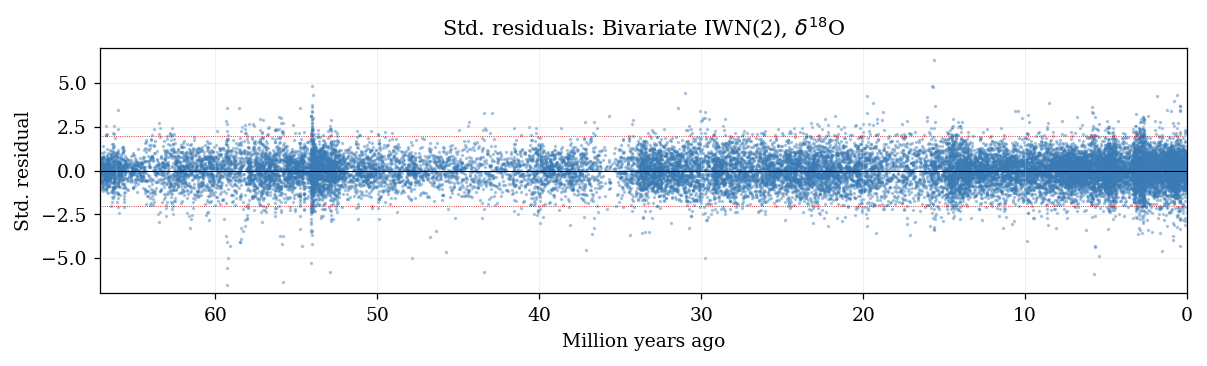}
\includegraphics[width=.45\textwidth]{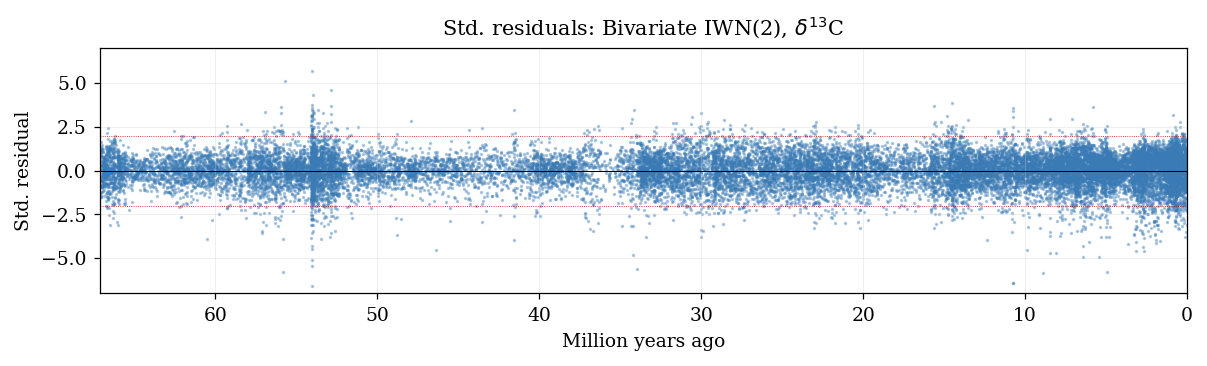}\\
\caption{\footnotesize Bivariate IWN($m=2$). Top row: Kalman-smoothed level (red) with 95\% pointwise confidence band (shaded) and data (blue dots) for $\delta^{18}$O (left) and $\delta^{13}$C (right). Bottom row: standardized prediction residuals for $\delta^{18}$O (left) and $\delta^{13}$C (right).\label{F:BivIWN2_d18O}}
\label{F:BivIWN2_d13C}
\end{figure}

\clearpage
\section{System matrices of the state space models}\label{A:Matrices}

\subsection{Univariate random walk plus noise specifications}

The measurement equation relates the data points from the data file of \cite{westerhold2020} to the unobserved components $\mu_t$ that are the central part of our modeling approach. A fact about the data set that has direct relevance for the dimension chosen for the measurement equation is that there are at most four data points at any given time stamp for each of the two time series, and they can originate from different studies. Denote the data points by $y$, the studies by $i$, the unobserved components by $\mu$, the climate states by $k$, and time by $t$.

The set of unique time stamps is given by $t_n$, $n\in\{1,\ldots, N\}$, $N=23722$. The time stamps $t_n$ fall into the interval  $[-67.10113,\, -0.000564]$, which is read as 67.10113 to 0.000564 million years ago. At time $t_n$,
\[
\underbrace{\left[\begin{array}{c}
y_{i_1,t_n} \\ y_{i_2,t_n} \\ y_{i_3,t_n} \\ y_{i_4,t_n} 
\end{array}\right]}_{y_{t_n}\in \R^{4\times 1}}
= \underbrace{\left[\begin{array}{c}
1 \\ 1 \\ 1 \\ 1
\end{array}\right]}_{Z \in \R^{4\times 1}}
\mu_{t_n} + 
\underbrace{\left[\begin{array}{c}
\eps_{i_1,t_n} \\ \eps_{i_2,t_n} \\ \eps_{i_3,t_n} \\ \eps_{i_4,t_n} 
\end{array}\right]}_{\eps_{t_n} \in \R^{4\times 1}},\; \mu_{t_n}\in\R.
\]
The index $i_{\{1,2,3,4\}} \in \{1,\,\ldots,\, 34\}$ denotes the 34 different studies of origin coded in the data file.  The zero-mean measurement equation error vector is assumed to be identically and independently normally distributed:
\[
\eps_{t_n} \stackrel{i.i.d.}{\sim} \mathsf{N}(0, H_{t_n}),
\]
with covariance matrix
\[
H_{t_n} = \text{diag}\left(\sigma^2_{\eps,i_1}, \sigma^2_{\eps,i_2}, \sigma^2_{\eps,i_3}, \sigma^2_{\eps,i_4}\right).
\]
The relation between $t_n$ and $i_{(\cdot)}$ is the mapping ``study $i_{(\cdot)}$ provided the data point $y_{i_\cdot,t_n}$ for time stamp $t_n$.''

The transition equation for the unobserved component is given as follows.
\[
\mu_{t_n+\Delta t_n} = \mu_{t_n} + \sqrt{\Delta t_n}\, \eta_{k, t_n},
\]
where $k\in \{1,\ldots, 6\}$ indicates the six different climate states identified in \cite{westerhold2020}. The zero-mean transition equation error is assumed to be identically and independently normally distributed: $\eta_{k, t_n}\stackrel{i.i.d.}{\sim}\mathsf{N}(0, Q_{t_n})$, $Q_{t_n} = \sigma_{\eta, k}^2$. The relation between $t_n$ and $k$ is the mapping ``time $t_n$ falls into period $k$ according to Table \ref{T:Climate_States}.''

\subsection{Bivariate random walk plus noise specifications}

The measurement equation is given as follows.
\[\arraycolsep=1.4pt\def\arraystretch{1.4}
\underbrace{\left[\begin{array}{c}
y_{i_1,t_n}^{\delta^{18}O} \\ y_{i_2,t_n}^{\delta^{18}O} \\ y_{i_3,t_n}^{\delta^{18}O} \\ y_{i_4,t_n}^{\delta^{18}O} \\
y_{i_1,t_n}^{\delta^{13}C} \\ y_{i_2,t_n}^{\delta^{13}C} \\ y_{i_3,t_n}^{\delta^{13}C} \\ y_{i_4,t_n}^{\delta^{13}C}
\end{array}\right]}_{y_{t_n}\in\R^{8\times 1}}
=
\underbrace{\left[\begin{array}{cc}
1 & 0 \\ 1 & 0 \\ 1 & 0 \\ 1 & 0 \\
0 & 1 \\ 0 & 1 \\ 0 & 1 \\ 0 & 1
\end{array}\right]}_{Z\in\R^{8\times 2}}
\underbrace{\left[\begin{array}{c}
\mu_{t_n}^{\delta^{18}O} \\ \mu_{t_n}^{\delta^{13}C}
\end{array}\right]}_{\mu_{t_n}\in\R^{2\times 1}}
+
\underbrace{\left[\begin{array}{c}
\eps_{i_1,t_n}^{\delta^{18}O} \\ \eps_{i_2,t_n}^{\delta^{18}O} \\ \eps_{i_3,t_n}^{\delta^{18}O} \\ \eps_{i_4,t_n}^{\delta^{18}O} \\
\eps_{i_1,t_n}^{\delta^{13}C} \\ \eps_{i_2,t_n}^{\delta^{13}C} \\ \eps_{i_3,t_n}^{\delta^{13}C} \\ \eps_{i_4,t_n}^{\delta^{13}C}
\end{array}\right]}_{\eps_{t_n}\in\R^{8\times 1}}.
\]
The distribution assumption is as above for the univariate model, simply extended to the longer error vector:
\[
H_{t_n} = \text{diag}\left(\sigma^2_{\eps,i_1,\delta^{18}O}, \sigma^2_{\eps,i_2,\delta^{18}O}, \sigma^2_{\eps,i_3,\delta^{18}O}, \sigma^2_{\eps,i_4,\delta^{18}O}, \sigma^2_{\eps,i_1,\delta^{13}C}, \sigma^2_{\eps,i_2,\delta^{13}C}, \sigma^2_{\eps,i_3,\delta^{13}C}, \sigma^2_{\eps,i_4,\delta^{13}C}\right),
\]
and the relation between $t_n$ and $i_{(\cdot)}$ is as before the mapping ``study $i_{(\cdot)}$ provided the data point $y_{i_\cdot,t_n}$ for time stamp $t_n$.''

The transition equation for the unobserved component $\mu$ is given as follows.
\[\arraycolsep=1.4pt\def\arraystretch{1.4}
\underbrace{\left[\begin{array}{c}
\mu_{t_n+\Delta t_n}^{\delta^{18}O} \\ \mu_{t_n + \Delta t_n}^{\delta^{13}C}
\end{array}\right]}_{\mu_{t_n+\Delta t_n}\in\R^{2\times 1}}
=
\underbrace{\left[\begin{array}{cc}
1 & 0 \\ 0 & 1
\end{array}\right]}_{T\in\R^{2\times 2}}
\underbrace{\left[\begin{array}{c}
\mu_{t_n}^{\delta^{18}O} \\ \mu_{t_n}^{\delta^{13}C}
\end{array}\right]}_{\mu_{t_n}\in\R^{2\times 1}}
+
\underbrace{\left[\begin{array}{c}
\eta_{k,t_n}^{\delta^{18}O}\\ \eta_{k,t_n}^{\delta^{13}C}
\end{array}\right]}_{\eta_{t_n}\in\R^{2\times 1}}.
\]
In the transition equation, we allow for correlation of the transition error from the unobserved component in $\delta^{18}$O with the one in $\delta^{13}$C: $\eta_{t_n}\stackrel{i.i.d.}{\sim}\mathsf{N}(0,Q_{t_n})$,
\[
Q_{t_n} = \left[\begin{array}{cc}
\sigma_{\eta,\delta^{18}O,k}^2 \Delta t_n^{\delta^{18}O} & \rho_k \sigma_{\eta,\delta^{18}O,k}\sigma_{\eta,\delta^{13}C,k} \min\{\Delta t_n^{\delta^{18}O}, \Delta t_n^{\delta^{13}C}\} \\
\rho_k\sigma_{\eta,\delta^{18}O,k}\sigma_{\eta,\delta^{13}C,k} \min\{\Delta t_n^{\delta^{18}O}, \Delta t_n^{\delta^{13}C}\} & \sigma_{\eta,\delta^{13}C,k}^2 \Delta t_n^{\delta^{13}C}
\end{array}\right],
\]
where $k$ indicates the climate state according to Table \ref{T:Climate_States}, and
\[
\Delta t_n^{(\delta^{18}O,\delta^{13}C)} = \left\{\begin{array}{l}
\Delta t_n,\qquad\qquad\qquad\rule{.6cm}{0cm}  \text{if preceding obs at $t_{n-1}$ is not NA},\\
\Delta t_n + \Delta t_{n-1}^{(\delta^{18}O,\delta^{13}C)},\quad \text{if preceding obs at $t_{n-1}$ is NA}.
\end{array}\right.
\]

\subsection{Bivariate random walk plus noise with Milankovitch forcing}\label{A:Milank}

The Milankovitch extension of Section~\ref{S:RWNBiv} adds a deterministic orbital term to the level transition. Let $e(t)$, $\varepsilon(t)$ and $p(t)$ denote eccentricity, obliquity and climatic precession from the astronomical solution La2004 \citep{laskar2004}, evaluated at observation age $t$, and write their increments over the interval to the next observation as $\Delta e_n = e(t_{n+1})-e(t_n)$, and analogously $\Delta\varepsilon_n$, $\Delta p_n$. The transition equation becomes
\[\arraycolsep=1.4pt\def\arraystretch{1.4}
\left[\begin{array}{c}
\mu_{t_n+\Delta t_n}^{\delta^{18}O} \\ \mu_{t_n+\Delta t_n}^{\delta^{13}C}
\end{array}\right]
=
\left[\begin{array}{cc} 1 & 0 \\ 0 & 1 \end{array}\right]
\left[\begin{array}{c}
\mu_{t_n}^{\delta^{18}O} \\ \mu_{t_n}^{\delta^{13}C}
\end{array}\right]
+
\underbrace{\left[\begin{array}{ccc}
b^{\delta^{18}O}_{e} & b^{\delta^{18}O}_{\varepsilon} & b^{\delta^{18}O}_{p} \\
b^{\delta^{13}C}_{e} & b^{\delta^{13}C}_{\varepsilon} & b^{\delta^{13}C}_{p}
\end{array}\right]}_{B\in\R^{2\times 3}}
\left[\begin{array}{c}
\Delta e_n \\ \Delta\varepsilon_n \\ \Delta p_n
\end{array}\right]
+
\left[\begin{array}{c}
\eta_{k,t_n}^{\delta^{18}O}\\ \eta_{k,t_n}^{\delta^{13}C}
\end{array}\right],
\]
with the state-disturbance covariance $Q_{t_n}$ unchanged. The six entries of $B$ are the coefficients reported in Table~\ref{T:Milank_Orb}. In the period-dependent variant of Table~\ref{T:Milank_OrbPer}, the matrix is climate-state specific, $B\to B_k$ with $k$ the climate state of Table~\ref{T:Climate_States}, so that there are six coefficient matrices and $6\times 6=36$ orbital coefficients. The orbital term is deterministic and so leaves the Kalman filter and smoother recursions otherwise unchanged.

\begin{figure}[htbp]
\centering
\includegraphics[width=\textwidth]{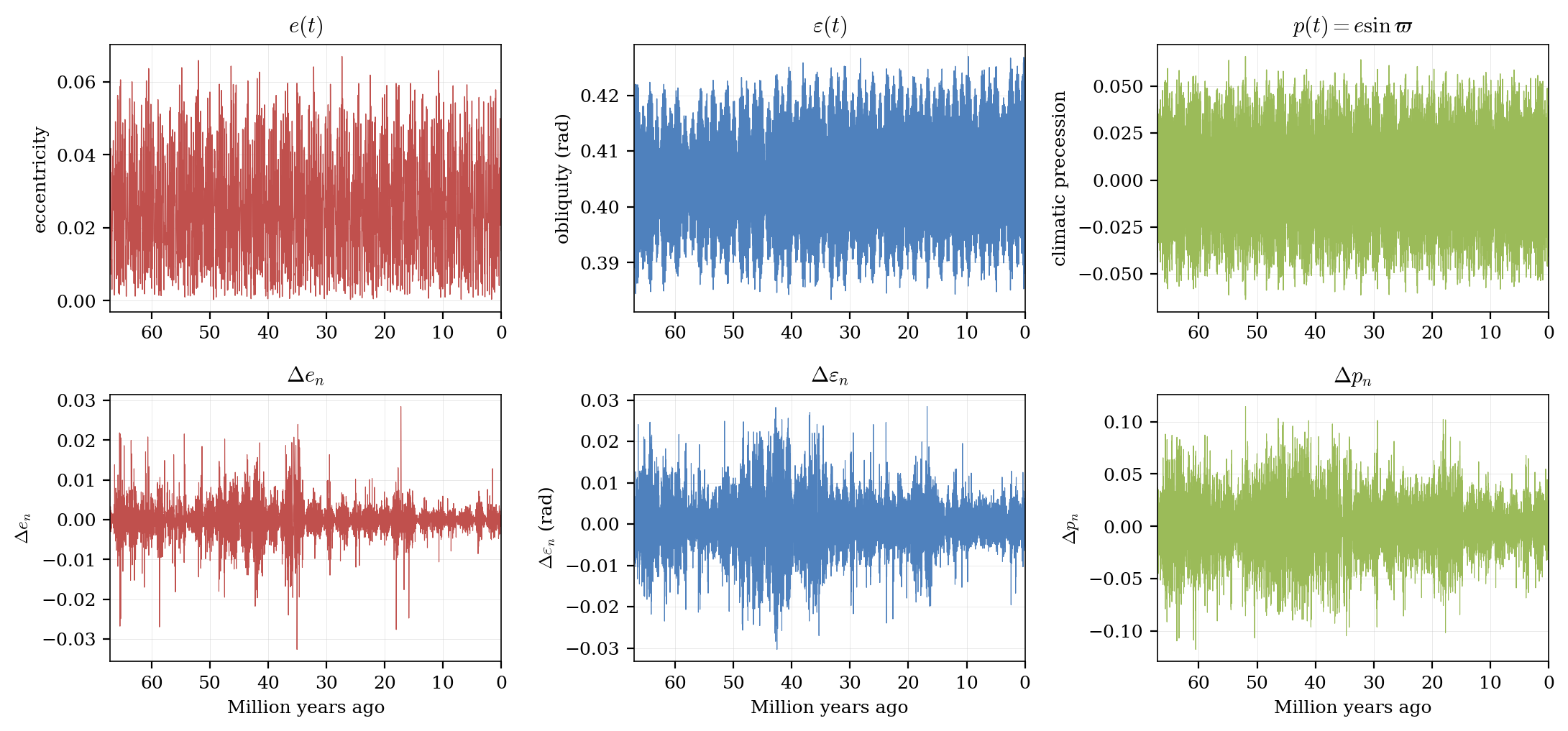}
\caption{\footnotesize Top row: the La2004 \citep{laskar2004} eccentricity $e(t)$, obliquity $\varepsilon(t)$ (in radians), and climatic precession $p(t)=e\sin\varpi$ over the $67$~Myr record span, evaluated on a fine time grid. Bottom row: their increments $\Delta e_n$, $\Delta\varepsilon_n$, $\Delta p_n$ over the inter-observation intervals, which are the deterministic regressors entering the level transition. The abscissa is age in millions of years before present.\label{F:Orbital}}
\end{figure}

Figure~\ref{F:Orbital} shows the three orbital series over the record span and the increments $\Delta e_n$, $\Delta\varepsilon_n$, $\Delta p_n$ that enter the transition as regressors. The top row displays the strong amplitude modulation of the La2004 solution (the eccentricity envelope, the changing amplitude of the obliquity cycles, and the precession amplitude tracking eccentricity through $p=e\sin\varpi$). The increments in the bottom row are larger in the early, sparsely sampled part of the record, where the inter-observation intervals $\Delta t_n$ are longer.

\clearpage

\section{Measurement variances: studies and benthic foraminifera species}\label{A:Sp}

\subsection*{Study variances}

The study-based specification differentiates the measurement-equation variance by the 34 studies of origin (column ``benthic Source'' in the data file of \cite{westerhold2020}) rather than by the ten drill sites used in the main text,
\begin{equation}
y_{n} = \mu_{t_n} + c_s + \eps_{n}, \qquad \eps_{n}\stackrel{indep.}{\sim} \mathsf{N}(0,\sigma_{\eps,i}^2),
\qquad i= i(n) \in \{1,\ldots,34\},\label{E:RWN_SV}
\end{equation}
and, combined with the climate-state transition variances and cross-proxy correlations of the preferred bivariate model~\eqref{E:BivCov} and study-specific constant intercepts $c_s$, gives the study counterpart of the bivariate site model in Table~\ref{T:RWN_SiteBiv}. Table~\ref{T:RWN_SPBiv} reports the estimates; the fit and the period transition variances and correlations are essentially the same as for the site-based specification in the main text. The estimated study measurement variances tend to decrease with the study's year of publication: a linear regression of the $\delta^{18}$O study variances on publication year gives a significantly negative slope of about $-0.001$ per year (adjusted $R^2=26\%$, $p=0.001$), and the $\delta^{13}$C regression a slope of about $-0.0026$ (adjusted $R^2=38\%$, $p<0.001$); both trends are essentially unchanged when the outlier ``McCarren et al.\ 2008'' is excluded (adjusted $R^2=28\%$ and $39\%$, respectively), suggesting that more recent measurements are on average more accurate.


\begin{table}
\caption{\footnotesize Maximum likelihood parameter estimates for model \eqref{E:BivCov} with measurement variances differentiated by the 34 studies of origin and study-specific constant intercepts $c_s$. Standard errors in parentheses. ``Barnet et al.\ 2017'' is the reference study ($c_{s_0}=0$); the study-specific intercepts $c_s$ are estimated jointly and omitted. ``Source'' refers to the entry in column ``benthic Source'' in tab ``Table S33'' of the data file \texttt{aba6853\_tables\_s8\_s34.xlsx} of \cite{westerhold2020}. The identifier ``McCarren et al. 2008 et al. 2008'' was merged with ``McCarren et al. 2008'' and ``Bickert et al.1997'' with ``Bickert et al. 1997''. The study Westerhold et al. 2020 is coded in the data file as ``this study''. BIC: Bayes information criterion.\label{T:RWN_SPBiv}}
\centering
\footnotesize
\begin{tabular}{lrcrc}\hline
Source                                          & No. obs.  & $\delta^{18}$O    & No. obs.  & $\delta^{13}$C \\\hline
$\sigma_{\eps,i}^2$ & & & & \\
``Barnet et al. 2017''                          & 548       & 0.0025 (0.0002)   & 548   & 0.0043 (0.0004)   \\
``Hull et al. 2020''                            & 29        & 0.0088 (0.0025)   & 29    & 0.0126 (0.0037)   \\
``Barnet et al. 2019''                          & 879       & 0.0064 (0.0004)   & 879   & 0.0028 (0.0002)   \\
``Littler et al. 2014''                         & 2164      & 0.0088 (0.0003)   & 2164  & 0.0053 (0.0003)   \\
``McCarren et al. 2008''                        & 47        & 0.0536 (0.0120)   & 47    & 0.0801 (0.0205)   \\
``Lauretano et al. 2015''                       & 636       & 0.0136 (0.0009)   & 636   & 0.0146 (0.0011)   \\
``Stap et al. 2010''                            & 372       & 0.0278 (0.0022)   & 372   & 0.0226 (0.0019)   \\
``Thomas et al. 2018''                          & 86        & 0.0136 (0.0024)   & 86    & 0.0151 (0.0027)   \\
``Lauretano et al. 2016; 2018''                 & 541       & 0.0054 (0.0005)   & 541   & 0.0048 (0.0006)   \\
``Sexton et al. 2011''                          & 301       & 0.0076 (0.0009)   & 301   & 0.0044 (0.0009)   \\
Westerhold et al. 2020                          & 1723      & 0.0077 (0.0004)   & 1723  & 0.0038 (0.0003)   \\
``Westerhold et al. 2018 LLTM''                 & 30        & 0.0112 (0.0033)   & 30    & 0.0456 (0.0130)   \\
``Boscolo Galazzo et al. 2014''                 & 34        & 0.0124 (0.0035)   & 34    & 0.0138 (0.0039)   \\
``Lear et al. 2004''                            & 182       & 0.0081 (0.0013)   & 182   & 0.0063 (0.0010)   \\
``Coxall et al. 2005''                          & 831       & 0.0090 (0.0006)   & 831   & 0.0094 (0.0006)   \\
``Coxall \& Wilson, 2011''                      & 413       & 0.0181 (0.0016)   & 413   & 0.0115 (0.0010)   \\
``Riesselmann et al. 2007''                     & 43        & 0.0167 (0.0042)   & 43    & 0.0307 (0.0077)   \\
``Paelike2006 PI Herrle''                       & 258       & 0.0202 (0.0022)   & 258   & 0.0128 (0.0015)   \\
``Wade and Paelike 2004''                       & 152       & 0.0424 (0.0058)   & 152   & 0.0268 (0.0037)   \\
``Liebrand et al. 2011;2016''                   & 3251      & 0.0109 (0.0004)   & 3251  & 0.0107 (0.0004)   \\
``Holbourn et al. 2015''                        & 892       & 0.0180 (0.0013)   & 892   & 0.0100 (0.0007)   \\
``Tian et al. 2014''                            & 61        & 0.0427 (0.0091)   & 61    & 0.0234 (0.0051)   \\
``Holbourn et al. 2014''                        & 1370      & 0.0134 (0.0007)   & 1370  & 0.0081 (0.0005)   \\
``Holbourn et al. 2007;2013;2018''              & 1987      & 0.0061 (0.0003)   & 1987  & 0.0065 (0.0003)   \\
``Drury et al. 2017''                           & 1129      & 0.0083 (0.0005)   & 1129  & 0.0172 (0.0009)   \\
``Tian et al. 2018''                            & 551       & 0.0102 (0.0008)   & 551   & 0.0253 (0.0018)   \\
UniCambridge Hodell / Westerhold et al. 2020    & 547       & 0.0150 (0.0011)   & 547   & 0.0103 (0.0008)   \\
``Bell et al. 2014''                            & 260       & 0.0071 (0.0008)   & 260   & 0.0162 (0.0017)   \\
MARUM Bremen / Westerhold et al. 2020           & 45        & 0.0091 (0.0021)   & 45    & 0.0127 (0.0030)   \\
``Franz \& Tiedemann2002''                      & 1304      & 0.0616 (0.0027)   & 1050  & 0.1085 (0.0051)   \\
``Tiedemann \& Franz 1997''                     & 330       & 0.0307 (0.0031)   & 264   & 0.0424 (0.0046)   \\
``Billups et al. 1998''                         & 538       & 0.0308 (0.0022)   & 538   & 0.0568 (0.0039)   \\
``Bickert et al. 1997''                         & 2562      & 0.0394 (0.0014)   & 2562  & 0.1293 (0.0041)   \\
``deMenocal et al. 1997''                       & 163       & 0.0138 (0.0025)   & 163   & 0.0743 (0.0097) \\\hline
$\sigma_{\eta,k}^2$                                 &           &                   &       &                           \\
67.10113 -- 56 Ma (``Warmhouse 2'')            & 2761      & 0.6102 (0.0537)   & 2761  & 0.9350 (0.0641)   \\
56 -- 47 Ma (``Hothouse'')                     & 3030      & 1.8654 (0.1128)   & 3030  & 3.3191 (0.1738)   \\
47 -- 34 Ma (``Warmhouse 1'')                  & 1786      & 0.6896 (0.0716)   & 1784  & 0.9534 (0.0828)   \\
34 -- 13.9 Ma (``Coolhouse 1'')                & 6669      & 1.8383 (0.1090)   & 6671  & 1.0801 (0.0727)   \\
13.9 -- 3.3 Ma (``Coolhouse 2'')               & 6282      & 1.7882 (0.1017)   & 6139  & 1.6792 (0.1031)   \\
3.3 -- 0.000564 Ma (``Icehouse'')              & 3731      & 11.5348 (0.7138)  & 3554  & 7.3876 (0.6943) \\\hline
$\rho_k$                                          &           &                   &       &                   \\
67.10113 -- 56 Ma (``Warmhouse 2'')            &   \multicolumn{4}{c}{0.8214 (0.0183)} \\
56 -- 47 Ma (``Hothouse'')                     &   \multicolumn{4}{c}{0.9604 (0.0048)} \\
47 -- 34 Ma (``Warmhouse 1'')                  &   \multicolumn{4}{c}{0.9111 (0.0117)} \\
34 -- 13.9 Ma (``Coolhouse 1'')                &   \multicolumn{4}{c}{0.5997 (0.0254)} \\
13.9 -- 3.3 Ma (``Coolhouse 2'')               &   \multicolumn{4}{c}{0.1261 (0.0362)} \\
3.3 -- 0.000564 Ma (``Icehouse'')              &   \multicolumn{4}{c}{-0.7889 (0.0333)} \\\hline
log-likelihood                                  &   \multicolumn{4}{c}{22760.72} \\
BIC                                             &   \multicolumn{4}{c}{-43990.16} \\\hline
\end{tabular}
\end{table}





\subsection*{Species variances}

The species-based specification instead differentiates the measurement variance by the 23 benthic foraminifera species (column ``benthic Species''), replacing $\sigma_{\eps,s}^2$ in \eqref{E:RWN_SitV} by
\begin{equation}
y_{n} = \mu_{t_n} + c_s + \eps_{n}, \qquad \eps_{n}\stackrel{indep.}{\sim} \mathsf{N}(0,\sigma_{\eps,j}^2),
\qquad j= j(n) \in \{1,\ldots,23\},\label{E:RWN_SpV}
\end{equation}
where $j$ indexes one of the 23 species. Estimating the preferred bivariate model~\eqref{E:BivCov} with these species-specific measurement variances and intercepts (and the Westerhold-period transition variances and correlations) gives the species counterpart of the bivariate study model; the estimates are reported in Table~\ref{T:BivSpV}.


\begin{table}
\caption{\footnotesize Maximum likelihood parameter estimates for model \eqref{E:BivCov} with measurement variances differentiated by the 23 raw benthic foraminifera species (column ``benthic Species'', no CSPP merge). Standard errors in parentheses. ``NTRUE'' is the reference species; the species-specific intercepts $c_s$ are estimated jointly and omitted. The species ``GORB'' and ``PMURR'' have no $\delta^{13}$C observations, so their $\delta^{13}$C parameters are not identified (``n/a''). BIC: Bayes information criterion.\label{T:BivSpV}}
\centering
\footnotesize
\begin{tabular}{lrcrc}\hline
Source & No. obs. & $\delta^{18}$O & No. obs. & $\delta^{13}$C \\\hline
$\sigma_{\eps,j}^2$ & & & & \\
NTRUE & 6949 & 0.0112 (0.0002) & 6949 & 0.0077 (0.0002) \\
OUMB & 240 & 0.0442 (0.0047) & 56 & 0.0482 (0.0104) \\
CSUBS or CEOCEA (unspecified) & 301 & 0.0074 (0.0008) & 301 & 0.0033 (0.0009) \\
CPRAE & 136 & 0.0185 (0.0026) & 136 & 0.0220 (0.0032) \\
CSPP & 2310 & 0.0165 (0.0006) & 2310 & 0.0296 (0.0011) \\
CSPP, whole specimen & 25 & 0.0247 (0.0091) & 25 & 0.0104 (0.0043) \\
CSPP, $>$250 & 353 & 0.0168 (0.0016) & 353 & 0.0079 (0.0008) \\
CSPP, $>$250, Reruns & 58 & 0.0115 (0.0026) & 58 & 0.0286 (0.0061) \\
CSPP, specimen $>$250 $\mu$m & 10 & 0.0001 (0.0000) & 10 & 0.0104 (0.0021) \\
CSPP, 150-250 & 2 & 0.0010 (0.0033) & 2 & 0.0002 (0.0005) \\
CGRIM & 318 & 0.0272 (0.0027) & 318 & 0.0196 (0.0018) \\
CHAVA & 92 & 0.0253 (0.0049) & 92 & 0.0120 (0.0024) \\
CMUND & 4380 & 0.0097 (0.0003) & 4380 & 0.0122 (0.0003) \\
PWUEL; CMUND & 4249 & 0.0099 (0.0004) & 4249 & 0.0076 (0.0003) \\
PWUEL (eitherCKULL or PWUEL) & 807 & 0.0125 (0.0008) & 807 & 0.0121 (0.0007) \\
CKULL & 255 & 0.0468 (0.0051) & 254 & 0.0475 (0.0055) \\
PWUEL & 3405 & 0.0433 (0.0013) & 3407 & 0.1214 (0.0033) \\
UVIG & 149 & 0.0485 (0.0064) & 133 & 0.0131 (0.0027) \\
GORB & 30 & 0.0552 (0.0169) & 0 & n/a \\
PMURR & 91 & 0.0640 (0.0106) & 0 & n/a \\
NUMB & 65 & 0.0590 (0.0126) & 65 & 0.1004 (0.0205) \\
CBRA & 22 & 0.0280 (0.0152) & 22 & 0.0739 (0.0285) \\
CCIC & 12 & 0.0594 (0.0319) & 12 & 0.0975 (0.0532) \\
\hline
$\sigma_{\eta,k}^2$ & & & & \\
67.10113 -- 56 Ma (``Warmhouse 2'') & 2761 & 0.3322 (0.0311) & 2761 & 0.5868 (0.0421) \\
56 -- 47 Ma (``Hothouse'') & 3030 & 2.3428 (0.1512) & 3030 & 4.2907 (0.2351) \\
47 -- 34 Ma (``Warmhouse 1'') & 1786 & 0.3139 (0.0365) & 1784 & 0.4647 (0.0432) \\
34 -- 13.9 Ma (``Coolhouse 1'') & 6669 & 2.3587 (0.1180) & 6671 & 1.1043 (0.0672) \\
13.9 -- 3.3 Ma (``Coolhouse 2'') & 6282 & 1.5967 (0.0931) & 6139 & 1.8126 (0.1036) \\
3.3 -- 0.000564 Ma (``Icehouse'') & 3731 & 11.8554 (0.7370) & 3554 & 8.3287 (0.7732) \\
\hline
$\rho_k$ & & & & \\
67.10113 -- 56 Ma (``Warmhouse 2'') & \multicolumn{4}{c}{+0.7830 (0.0253)} \\
56 -- 47 Ma (``Hothouse'') & \multicolumn{4}{c}{+0.9687 (0.0039)} \\
47 -- 34 Ma (``Warmhouse 1'') & \multicolumn{4}{c}{+0.8423 (0.0223)} \\
34 -- 13.9 Ma (``Coolhouse 1'') & \multicolumn{4}{c}{+0.6440 (0.0224)} \\
13.9 -- 3.3 Ma (``Coolhouse 2'') & \multicolumn{4}{c}{+0.0920 (0.0377)} \\
3.3 -- 0.000564 Ma (``Icehouse'') & \multicolumn{4}{c}{-0.7257 (0.0361)} \\
\hline
log-likelihood & \multicolumn{4}{c}{20967.97} \\
BIC & \multicolumn{4}{c}{-40847.94} \\\hline
\end{tabular}
\end{table}

\clearpage


\clearpage

\bibliographystyle{sn-chicago}
\bibliography{bhk_paleo_v13_cd}

\clearpage

\section*{Acknowledgements}

For helpful comments and suggestions, we thank participants at the conferences on Econometric Models of Climate Change (EMCC-VII and VIII) in Amsterdam in 2023 and in Cambridge, UK, in 2024, at the General Assembly of the EGU in Vienna in 2024, and participants at the Earth Day Hackathon at Aalborg University.

\end{document}